\documentstyle[eqsecnum,preprint,aps]{revtex}
\newcommand{\be}{\begin{equation}}
\newcommand{\ee}{\end{equation}}
\newcommand{\bea}{\begin{eqnarray}}
\newcommand{\eea}{\end{eqnarray}}
\newcommand{\OP}{\vec{\psi}}
\newcommand{\SG}{\vec{\sigma}}
\newcommand{\M}{\vec{m}}
\newcommand{\U}{\vec{u}}
\newcommand{\F}{{\cal F}}
\newcommand{\A}{\omega_{0}}
\newcommand{\B}{\omega_{2}}
\newcommand{\C}{\omega_{1}}
\newcommand{\T}{\vec{\Theta}}
\begin{document}
\draft
\title{Phase ordering of the $O(2)$ model in the post-gaussian approximation}
\author{Robert A. Wickham and Gene F. Mazenko}
\address{The James Franck Institute and the Department of Physics\\ The
University of Chicago\\ Chicago, Illinois 60637}
\date{\today}
\maketitle
%
%
\begin{abstract}
The gaussian closure approximation previously used to study the
growth kinetics of the non-conserved $O(n)$ model is shown to be the
zeroth-order approximation in a well-defined sequence of approximations
composing a
more elaborate theory. This paper studies the effects of including the next
non-trivial correction in this sequence for the case $n=2$. The scaling forms
for the order-parameter and order-parameter squared correlation
functions are determined for the physically interesting cases of the
$O(2)$ model in two and three spatial dimensions. The post-gaussian
versions of these quantities show improved agreement with simulations.
Post-gaussian formulae for the defect density and the defect-defect
correlation function $\tilde{g}(x)$ are derived. As in the previous gaussian
theory, the addition
of fluctuations allows one to eliminate the unphysical divergence in
$\tilde{g}(x)$ at short scaled-distances. The non-trivial exponent $\lambda$,
 governing the decay of order-parameter auto-correlations, is computed in this
approximation both with and without fluctuations.
\end{abstract}
%
%
\section{Introduction}

In the gaussian closure approximation for phase-ordering kinetics
the order-parameter
field is expressed in terms of an auxiliary field which is assumed to
obey gaussian statistics \cite{MAZENKO90,LIU92a}.
Recently, one of the authors extended this approach
to treat more general statistics and applied the method successfully to the
cases of conserved and non-conserved scalar fields
\cite{MAZENKO94a,MAZENKO94b}. This post-gaussian approach successfully
eliminated several shortcomings of the gaussian theory.
The present paper has two primary goals. The first goal is to generalize the
post-gaussian theory to treat the non-conserved $O(n)$ symmetric
model with  continuous symmetry $(n>1)$. Systems with continuous
symmetry have many physical realizations, including ferromagnets,
superfluids, and liquid crystals \cite{CHUANG93}. We will focus on the case
$n=2$, where the defect-defect correlation function, $\tilde{g}(x)$, has an
unphysical divergence at short scaled-distances, $x$, in the gaussian theory.
While the post-gaussian theory weakens this divergence, it is not eliminated.
The second goal of this paper is then to show that the inclusion of
fluctuations counter-balance and thereby eliminate the divergence
in $\tilde{g}(x)$ in the post-gaussian case.

In a phase-ordering scenario, the dynamical evolution of the order parameter
$\OP(1) = \OP ({\bf r}_{1},t_{1})$ is not typically governed
by a gaussian probability distribution, and analytical progress up to now has
relied on relating $\OP(1)$ to an auxiliary field $\M(1)$,
assumed to be gaussian. The gaussian approximation has been very
successful in treating the scaling inherent in the late-time behaviour of
a host of growth kinetics problems \cite{QSTATE}.
The gaussian theory predicts the now well-established result that for late
times following a quench from the disordered to the ordered phase the
dynamics obey scaling and the system can be described in
terms of a single growing length
%
\be
\label{EQ:GROWTHLAW}
L(t) \sim t^{\phi},
\ee
which is characteristic of the spacing between defects at time $t$ after the
quench. $\phi$ is a growth exponent. In this regime the
scaling form $\F$ for the equal-time ($t_{1} = t_{2} = t$)
order-parameter correlation function
%
%
\bea
\label{EQ:OPCOR}
C_{\psi}(12) & \equiv & \langle \OP (1) \cdot \OP (2) \rangle \\
& = & \psi_{0}^{2} \F (x) \nonumber
\eea
can be accurately calculated within the theory.
Here $\psi_{0}$ is the magnitude $\psi = |\OP|$ of the order-parameter in the
ordered phase.  The scaled length $x$ is defined as $x = r/L(t)$ with
$r \equiv |{\bf r}| \equiv |{\bf r}_{2} - {\bf r}_{1}|$.
For the non-conserved models considered here the growth exponent is
$\phi = 1/2$,
which is predicted by the theory and well established by experiments and
simulations \cite{N=2D=2}. The gaussian theory also makes quantitatively
accurate
predictions \cite{LIU92a,LIU91,LEE95} for the exponent $\lambda$
governing the decay
of the order-parameter auto-correlations, and defined by
%
\be
\label{EQ:LAMDEF}
C_{\psi}({\bf 0},t,t') \sim \frac{1}{L^{\lambda}(t)} \mbox{ for } t \gg t'.
\ee
Finally, in addition to these accomplishments, the
gaussian approximation is relatively easy to implement, and has
straightforward generalizations to more complex systems.

Despite these achievements there remain a few unresolved issues. The
approximate nature of gaussian statistics was highlighted in the work
of Blundell, Bray, and Sattler (BBS) \cite{BLUNDELL93,BLUNDELL94}
where they computed, within the gaussian closure approximation,
the two-point correlation function for the square of the order-parameter
field:
%
\be
C_{\psi^{2}}(12) = \frac{\langle [\psi_{0}^{2} - \psi^{2}(1)][\psi_{0}^{2} -
 \psi^{2}(2)] \rangle}{\langle \psi_{0}^{2} - \psi^{2}(1) \rangle \langle
 \psi_{0}^{2} - \psi^{2}(2) \rangle} - 1.
\label{EQ:CPSI2}
\ee
It is usual in comparing the theoretical scaling function $\F(x)$ with
the data (both of which are relatively featureless) to rescale the
length $x$
to give the best fit. This rescaling reflects the uncertainty in the relation
of the theoretical pre-factor of the power law growth (\ref{EQ:GROWTHLAW})
to the pre-factor
determined in experiment and simulation \cite{PREFACTOR}.
By plotting $1/(C_{\psi^{2}}+1)$ against
 $1 - \F$, hereafter referred to as a BBS plot, BBS were able to eliminate
any adjustable
fitting parameter and show that there were qualitative differences between the
simulation results \cite{BLUNDELL94} and the predictions of existing gaussian
theories (see Fig. \ref{FIG:BBS}).
 These discrepancies indicated a need to go beyond the
gaussian approximation. Another motivation for going beyond the gaussian
approximation is the {\em ad hoc} nature of the approximation itself and the
desire to encompass it within a more general and systematic framework
\cite{BRAY93}.

A second problem with the gaussian closure
approximation occurs when one examines
defect correlations. In this paper we focus on point defects ($n=d$) whose
density
is defined as
%
%
\be
\label{EQ:DEFDENSITY}
\rho(1) = \sum_{\alpha} q_{\alpha} \mbox{ }
          \delta({\bf r}_{1} - {\bf x_{\alpha}}(t_{1}))
\ee
where ${\bf x_{\alpha}}(t_{1})$ is the position at time $t_{1}$ of the
$\alpha$th point defect, which has a topological charge $q_{\alpha}$.
Defect-defect correlations
%
%
\be
\label{EQ:DEFDENCOR}
{\cal G}(12) \equiv \langle \rho (1) \rho(2) \rangle
\ee
at equal-times can be shown \cite{LIU92b} to decompose into
two parts
%
%
\be
\label{EQ:POINTDEF}
{\cal G}({\bf r}, t) = n_{0}(t) \delta({\bf r}) + g_{dd}({\bf r},t).
\ee
The quantity $n_{0}(t)$ represents defect self-correlations and is just
the total unsigned number density of defects at time $t$. We will be
primarily concerned here with the defect-defect correlation function
$g_{dd}({\bf r},t)$
which measures the correlations between different defects. The conservation
law
%
\be
\label{EQ:CONSERVATION}
\int d^{n} r \mbox{ } {\cal G} ({\bf r},t) = 0
\ee
relates $n_{0}(t)$ and $g_{dd}({\bf r},t)$ through
%
\be
\label{EQ:N0GDD}
n_{0}(t) = - \int d^{n} r \mbox{ } g_{dd}({\bf r},t).
\ee
In the
scaling regime it can be shown \cite{LIU92b} that $n_{0}(t)
\sim L^{-n}(t)$ and that
$g_{dd}({\bf r},t)$ has the form
%
%
\be
\label{EQ:DEFDEF}
g_{dd}({\bf r},t) = \frac{1}{L^{2n}(t)} \tilde{g}(x)
\ee
where $\tilde{g}(x)$ is a universal scaling function. While
the form for $\tilde{g}(x)$ obtained from the gaussian theory for the
two-dimensional $O(2)$ model is in good agreement with simulations
\cite{MONDELLO90} and experiments \cite{NAGAYA95} at
large scaled-distances, there are qualitative differences in the
short scaled-distance behaviour. The gaussian theory \cite{LIU92b}
predicts a divergence in $\tilde{g}(x)$ at small $x$ while experiments and
simulations have $\tilde{g}(x)$ approaching zero at the origin
(see Fig. \ref{FIG:DEFECT}).

This paper addresses these issues for the non-conserved $O(n)$ model by using
a well-defined sequence of approximations for the probability
distribution of the auxiliary field $\M$, which reduces to a gaussian
distribution at lowest order. The theory presented here treats in detail the
next non-trivial term in the sequence. This post-gaussian approach has been
successfully applied to both the cases of a non-conserved \cite{MAZENKO94a}
and a conserved \cite{MAZENKO94b} scalar order parameter. While, as expected,
the non-conserved scalar theory predicts little change in the form for
$\F$, the BBS plot shows a marked improvement over the gaussian theory
when compared
with simulations. This encourages one extend the post-gaussian theory to the
$O(n)$ case. The key result of this paper is that $\F$,
$C_{\psi^{2}}$, $n_{0}(t)$, $\tilde{g}(x)$ and $\lambda$ can all be extracted
using non-gaussian statistics. This is non-trivial since these quantities have
universal forms in the scaling regime. This is connected to the fact that the
probability distribution governing the auxiliary field in the gaussian case
has a fixed-point
form determined by the solution of an eigenvalue problem.
In the post-gaussian case the determination of the fixed-point form for the
probability distribution
requires the solution of a double eigenvalue problem.
This paper focuses on the $O(2)$ model in two and three spatial dimensions
where experimental and simulation results are readily available.
For the $O(2)$
model in the post-gaussian approximation we find that the form for $\F$
changes little from the gaussian results and, as in the scalar case, the
agreement of the BBS plot with simulations is improved. This improvement
includes a dependence on dimensionality
seen in the simulations, but not exhibited by the gaussian theory. There
are some problems related to the negativity of $C_{\psi^{2}}$ at small $x$ but
these are seen to be a consequence of the manner in which we defined the
sequence of approximations. The exponent $\lambda$ is now in poorer agreement
with the simulation data.
The divergence in $\tilde{g}(x)$ is found to
be weaker in the post-gaussian theory than in the gaussian theory, but it is
not completely eliminated.

We recently showed that this divergence can be eliminated in the gaussian case
\cite{MAZENKO96} if one includes the effects of fluctuations
about the ordering field. The origin of the divergence is the
appearance of non-analytic terms in the small-$x$ expansion of the auxiliary
field correlation function $f$. One
removes the divergence by eliminating the non-analytic terms in $f$ through
a careful treatment of the fluctuations. This development is theoretically
pleasing since
one expects the auxiliary field correlation function to be well-behaved.
The question remains whether this post-gaussian scheme can be
smoothly generalized to include these fluctuations. We answer this question
in the affirmative and see that the post-gaussian theory with fluctuations
is a rather natural generalization of the gaussian theory. Again, the
divergence
in $\tilde{g}(x)$ is removed and the magnitude of $\tilde{g}(0)$ is
reduced, bringing the post-gaussian theory into better
agreement with the simulation results \cite{MONDELLO90} than the gaussian
theory (see Fig. \ref{FIG:DEFECT}).

The first part of this paper is mainly devoted to developing the results for
the post-gaussian theory. Later, after we present the post-gaussian results
for $\tilde{g}(x)$, we will discuss the role of fluctuations in detail.
 Section II
reviews the gaussian $O(n)$ model and the separation of the equation of motion
into an equation for the ordering field and an equation for the fluctuations.
The main results of this paper are contained in Section III which, after
introducing the post-gaussian probability distribution and general formulae
for calculating post-gaussian averages, derives post-gaussian
expressions for $\F$, $C_{\psi^{2}}$ and the equations of motion. Defect
correlations are discussed, leading into Section IV which considers
the inclusion of fluctuations in the post-gaussian
theory. The results of numerical analysis of the new, post-gaussian
eigenvalue problem are presented in Section V. We conclude with a summary and
discussion.
%
%
\section{Model}
%
%
\subsection{Preliminaries}

We consider the $O(n)$ model, which describes the dynamics
of a  non-conserved, $n$-component order-parameter field $\OP (1) =
(\psi_{1} (1), \cdots,\psi_{n} (1) )$. To begin we will work with general $n$;
however, later we will focus on the interesting case $n=2$.
As in previous work in this area \cite{LIU92a}, the dynamics are
modeled using a time-dependent Ginzburg-Landau equation
%
\be
\frac{\partial \OP}{\partial t} = - \Gamma \frac{\delta F[\OP]}
{\delta \OP}.
\label{EQ:LANG}
\ee
We assume that the quench is to zero temperature where the usual noise
term on the right-hand side is zero \cite{BRAY89}.
 $\Gamma$ is a kinetic coefficient and
$F[\OP]$ is the free-energy, assumed to be of the form
%
\be
\label{EQ:FREEENERGY}
F[\OP] = \int d^{d} r ( \frac{c}{2} |\nabla \OP|^{2} + V[\psi])
\ee
where the potential $V[\psi]$ is chosen to have $O(n)$ symmetry and a
degenerate ground state with $|\OP| = \psi_{0}$ \cite{POTENTIAL}.
With a suitable redefinition of the time and
space scales the coefficients $\Gamma$ and $c$ can be set to one and
(\ref{EQ:LANG}) can be written as
%
\be
\frac{\partial \OP}{\partial t} = \nabla^{2} \OP -
\frac{\partial V[\psi]}{\partial \OP}.
\label{EQ:MOT}
\ee
It is believed that our final results are independent of the exact nature of
the initial state, provided it is a disordered state.

The evolution induced by
(\ref{EQ:MOT}) causes $\OP$ to order and assume a distribution that is far
from gaussian. It is by now standard to introduce a mapping between the
physical field $\OP$ and an auxiliary field $\M$ with more
tractable statistics. We can decompose $\OP$ exactly as
%
\be
\label{EQ:MAPPING}
\OP = \SG [ \M ] + \U.
\ee
The utility of this decomposition lies in our ability to create a
consistent theory with the mapping $\SG$ chosen to reflect the defect
structure in the problem, and the fluctuation correction
$\U$ constructed to be small at late-times.
The precise statistics satisfied by the fields $\M$ and $\U$  will be
specified below.

The mapping giving $\SG$ as a function of $\M$ is chosen to incorporate the
dominant defect structure in the late-stage ordering kinetics. We assume that
$\SG$ satisfies the
Euler-Lagrange equation for the free-energy given by (\ref{EQ:FREEENERGY})
with the spatial coordinate replaced by the auxiliary field:
%
\be
\nabla^{2}_{m} \SG [\M] = \frac{\partial V [\SG]}{\partial \SG }.
\label{EQ:SIGPOT}
\ee
The defects are then the non-uniform solutions of (\ref{EQ:SIGPOT}) which
match on to the uniform solution at infinity. Since we expect only the
lowest-energy defects, having unit topological charge, will survive to
late-times the relevant solutions to (\ref{EQ:SIGPOT}) will be of the form
%
\be
\SG [\M] = A(m) \hat{m}
\label{EQ:SIGEX}
\ee
where $m = |\M| \mbox{ and } \hat{m} = \M/m$. Thus the interpretation of
$\M$ is that its magnitude represents the distance away from a defect core and
its orientation indicates the direction to the defect core. We expect
$m$, away from the defect cores, to grow as $L$ in the
late-time scaling regime. Inserting (\ref{EQ:SIGEX}) into (\ref{EQ:SIGPOT})
gives an equation for $A$,
%
\be
\nabla^{2}_{m} A - \frac{n-1}{m^{2}} A - V'[A] = 0,
\label{EQ:A}
\ee
where the prime indicates a derivative with respect to $A$.
The boundary conditions are $A(0) = 0, \mbox{ } A(\infty) = \psi_{0}$. An
analysis of (\ref{EQ:A}) for $n>1$ and large $m$ yields
%
\be
A(m) = \psi_{0} \left[ 1 - \frac{\kappa}{m^{2}} + \cdots \right]
\label{EQ:A-LGM}
\ee
where $\kappa = (n-1)/V''[\psi_{0}] > 0$.
The algebraic relaxation of the order-parameter to its ordered value is a
distinct feature of the $O(n)$ model for $n>1$ which must be carefully
treated in the evaluation of some averages. In the scalar case $(n=1)$
$\psi$ relaxes exponentially to $\psi_{0}$ away from the defects.
%
%
\subsection{Separation of the equations of motion}

In \cite{MAZENKO96} it was shown how one can quite generally separate the
equation of motion (\ref{EQ:MOT}) into an equation for the ordering field
$\SG$ and an equation for the fluctuating field $\U$. One finds
that $\SG$ satisfies the equation of motion
%
%
\be
\label{EQ:SIGMAMOT}
\vec{B} = 0
\ee
with
%
\be
\label{EQ:BDEF}
\vec{B} \equiv \partial_{t} \SG  - \nabla^{2} \SG
+ \nabla_{m}^{2} \SG - \T.
\ee
$\T$ is chosen so that $\U$ is small in the scaling regime and thus
represents a fluctuation. The equation of motion for $\U$ is
%
%
\be
\label{EQ:UONLY}
\frac{\partial u_{i}}{\partial t} = \nabla^{2}u_{i}
-W_{ij}[\SG] u_{j} - \Theta_{i}
\ee
where a sum over the index $j$ is assumed and, to leading order in $1/L$,
%
\be
W_{ij}[\SG ] =q_{0}^{2}\hat{\sigma}_{i}\hat{\sigma}_{j}
\ee
which is purely longitudinal.  $q_{0}^{2} = V''[\psi_{0}] > 0$.

If we set $\vec{\Theta}$ equal to zero in (\ref{EQ:BDEF}) we obtain the
equation used previously to determine the $\SG$ correlations \cite{LIU92a}.
This choice decouples $\SG$ and $\U$. The equation for $\U$ would then
separate into a (massless \cite{MASS}) diffusion equation for the transverse
piece $\U_{T}$ and
an equation for the longitudinal piece $u_{L}$ with a mass term
$-q_{0}^{2} u_{L}$.
However, as was seen in the gaussian theory \cite{LIU92a,LIU92b}, the
equation for $\SG$ would necessarily lead to non-analytic
behaviour in the short scaled-distance expansion for the normalized auxiliary
 field correlation function, $f(x)$, and would ultimately lead to an unphysical
divergence in $\tilde{g}(x)$ at small $x$. We must
choose $\vec{\Theta}$ so that our auxiliary field (or fields, in the
post-gaussian theory) is analytic for
small $x$.  The possible forms we can use for $\vec{\Theta}$ are discussed
in detail in \cite{MAZENKO96}. The key ideas are that

(i) $\vec{\Theta}$ is chosen to be a function of $\M$ only.  This choice
    means that $\SG$ satisfies a closed equation, while $\U$
    is slaved by $\M$. This highlights the fact that, since we are working
    at zero temperature, it is $\T$ and not temperature which is driving
    the fluctuations \cite{KRAMER94}.

(ii)  $\vec{\Theta}$ must be odd under $\M \rightarrow -\M$.

(iii)  $\vec{\Theta}$ must scale as $ {\cal O}(L^{-2})$ in the
scaling regime if it is to compensate for the terms in the
equation of motion which lead to the non-analyticities in
$f$. This will also allow us to treat $\vec{u}$ as a fluctuation
since it will imply $\vec{u} \sim L^{-2}$.

It is sufficient for our purposes to consider
%
\be
\label{EQ:TFORM}
\T = \frac{\A}{L^{2}(t)} \SG,
\ee
where $\A$ is a constant to be determined.
This form will allow us to  construct $f(x)$ to be analytic through
terms of ${\cal O} (x^{2})$.
It was shown in \cite{MAZENKO96} how (\ref{EQ:TFORM}) is the first term is
a series for $\T$
which can be used to enforce analyticity at ${\cal O} (x^{4})$ and beyond.
%
%
\subsection{The gaussian approximation}

To complete the definition of the model one must specify the form of
the probability distribution for the auxiliary field $\M$. Forcing
$\SG$ to satisfy the exact equation of motion (\ref{EQ:SIGMAMOT})
is tantamount to solving the problem exactly, and will determine a
probability distribution for $\M$ which is complicated and extremely
difficult for purposes of computation. Progress can be made if one
imposes the weaker constraint
%
\be
\label{EQ:TWOPTCONST}
\langle \vec{B}(1) \cdot \SG (2) \rangle_{0} = 0.
\ee
This equation allows one to insure that $\vec{B}(1)$ is reasonably small at
late-times but gives one the flexibility to choose a suitable probability
distribution. The simplest choice is a gaussian probability distribution for
$\M$:
%
\be
P[\M] = P_{0}[\M] \equiv {\cal N} e^{-K_{0} [\M]}
\ee
with ${\cal N}$ a normalization constant and
%
\be
K_{0}[\M] = \frac{1}{2} \int d1d2 \mbox{ } C_{0}^{-1}(12) \M (1) \cdot \M (2).
\ee
$C_{0} (12)$, the correlation function for the auxiliary field, is explicitly
defined through
%
\be
\langle m_{i} (1) m_{j} (2) \rangle_{0} = \delta_{ij} \mbox{ } C_{0} (12).
\ee
Here we have used $\langle \cdots \rangle_{0}$ to indicate an average
using the gaussian distribution $P_{0}[\M]$. Later, when we consider
post-gaussian
statistics, we will use $\langle \cdots \rangle$ to denote an average using
the post-gaussian distribution $P[\M]$.
The system is assumed to be statistically isotropic and homogeneous so
$C_{0}(12)$ is invariant under interchange of its spatial indices. For future
reference we also define the one-point correlation function
%
%
\be
S_{0}(1) = C_{0}(11)
\ee
and the normalized auxiliary field correlation function
%
%
\be
\label{EQ:DEFNF}
f(12) = \frac{C_{0}(12)}{\bar{S}_{0}(12)}
\ee
with $\bar{S}_{0}(12) = \sqrt{S_{0}(1)S_{0}(2)}$. As discussed above it is
expected that both $C_{0}$ and $S_{0}$ grow as $L^2$ at late times.

In the gaussian theory the  relationship between the auxiliary function $f$
and the observable
functions $C_{\psi}$, $C_{\psi^{2}}$
and $\tilde{g}$ can be derived without reference to the dynamics contained in
(\ref{EQ:TWOPTCONST}).
Using (\ref{EQ:MAPPING}), (\ref{EQ:SIGEX}) and (\ref{EQ:A-LGM})
$C_{\psi}$ can be written to leading order in $1/L$ as
%
%
\bea
\label{EQ:OPCOR2}
C_{\psi}(12) & = & \psi_{0}^{2} \langle \hat{m} (1) \cdot \hat{m} (2)
\rangle_{0} \\
\nonumber     & = & \psi_{0}^{2} \F(12),
\eea
with
%
%
\be
\label{EQ:HYPER}
\F = \frac{nf}{2 \pi} B^{2} \left[ \frac{1}{2},\frac{n+1}{2}
\right]
F \left[ \frac{1}{2}, \frac{1}{2};\frac{n+2}{2};f^{2} \right]
\ee
where $B$ is the beta function and $F$ is the hypergeometric function
\cite{LIU92a,BRAY91}.
For later convenience, we define the short-forms
%
\be
F_{1} \equiv  F \left[ \frac{1}{2}, \frac{1}{2};\frac{n+2}{2};f^{2} \right]
\ee
and denote
\be
F'_{1} \equiv \frac{d F_{1}}{df} \mbox{ , } F''_{1} \equiv \frac{d^{2} F_{1}}
{df^{2}}, \dots
\ee

The quantity $C_{\psi^{2}}$ (\ref{EQ:CPSI2}), as
 obtained by BBS \cite{BLUNDELL93} for $n>1$, is given by
%
\be
C_{\psi^{2}} = F \left[ 1, 1; \frac{n}{2}; f^{2} \right] - 1.
\ee
For later notational simplicity we shall write
%
\be
F_{2} =  F \left[ 1, 1; \frac{n}{2}; f^{2} \right]
\ee
and, as with $F_{1}$, use the prime notation for differentiation with
respect to $f$. As pointed out in  \cite{BLUNDELL94},
one can directly test the assumption of gaussian statistics, independent
of the  spatial form of $f$, by plotting
$1/(C_{\psi^{2}}+1)$ against $1 - \F$. This is done for the gaussian $O(2)$
model in Fig. \ref{FIG:BBS} and the discrepancy
between the gaussian theory and the simulation data is evident.
Since this discrepancy is due to the choice of gaussian statistics and not
due to the method  used to determine $f$ it strongly suggests that an
improved choice of probability distribution is needed.

Within the gaussian theory, $\tilde{g}(x)$ is given by \cite{LIU92b}
%
%
\be
\label{EQ:GDEFECTDEFECTCOR}
\tilde{g} (x) = n! \left( \frac{h}{x} \right)^{n-1}
\frac{\partial h}{\partial x}
\ee
with
%
\be
\label{EQ:HDEF}
h = - \frac{\gamma}{2 \pi} \frac{\partial f}{\partial x}
\ee
and $\gamma = 1/\sqrt{1-f^2}$. The defect density is
%
%
\be
\label{EQ:GDEFDEN}
n_{0}(t) = \frac{n!}{2^{n} \pi^{n/2} \Gamma(1 + n/2)} \left[
\frac{S_{0}^{(2)}}{n S_{0}(t)} \right]^{n/2}
\ee
with
%
%
\be
\label{EQ:DEFS02}
S_{0}^{(2)} = \frac{1}{n} \langle [\vec{\nabla} \M ]^{2} \rangle_{0}.
\ee
In the gaussian theory without fluctuations $S_{0}^{(2)} = 1$.
The expressions (\ref{EQ:GDEFECTDEFECTCOR}) and
(\ref{EQ:GDEFDEN}) satisfy the conservation law (\ref{EQ:N0GDD}).

The constraint (\ref{EQ:TWOPTCONST}) determines the time evolution of the
two-point order-parameter correlations. We use (\ref{EQ:TWOPTCONST}) to
determine $f$ and $\F$, which are related through (\ref{EQ:HYPER}).
Knowledge of $f$
allows us then to determine $C_{\psi^{2}}$ and $\tilde{g}$. To simplify the
discussion we restrict ourselves initially to the case $\T = 0$, considered in
\cite{LIU92a}. The case when $\T$ has the form (\ref{EQ:TFORM}) was addressed
in \cite{MAZENKO96}, and later we will discuss the inclusion
of fluctuations in some detail. After some manipulation (\ref{EQ:TWOPTCONST})
becomes
%
\be
\partial_{t_{1}} \F (12) - \nabla_{1}^{2} \F (12) - \frac{1}{S_{0}(1)}
f \partial_{f} \F (12) = 0.
\label{EQ:G-EQNF}
\ee
Here we have used the short-hand notation $\partial_{f} \F =
\partial \F/\partial f$. For $t_{1} \gg t_{2}$ both $\F$
and $f$  are small. In this limit (\ref{EQ:G-EQNF}) becomes a linear equation
for $\F$ and, with the definition
%
\be
\label{EQ:GLENGTH}
L^{2} (t) = \frac{\pi S_{0} (t)}{2 \mu} =  4 t
\ee
for the scaling length $L$ \cite{MU}, $\lambda$ can be determined as
\cite{LIU92a,LIU91}:
%
\be
\label{EQ:GLAMBDA}
\lambda = d - \frac{\pi}{4 \mu}.
\ee
To examine the equal-time order-parameter correlations in the late-time
scaling regime we set $t_{1}=t_{2}=t$ and write (\ref{EQ:G-EQNF}) in terms
of the scaled distance $x$. To leading order in $1/L$ we have \cite{LIU92a}
%
%
\be
\label{EQ:G-SCALE}
\vec{x} \cdot \nabla_{x} \F + \nabla_{x}^{2} \F +
\frac{\pi}{2 \mu} f \partial_{f} \F = 0.
\ee
This is a non-linear eigenvalue problem with eigenvalue $\mu$, entering via
the definition (\ref{EQ:GLENGTH}) for the scaling length. $\mu$ is determined
by numerically matching the analytically determined behaviour at small-$x$
onto the analytically determined behaviour at large-$x$.
For large $x$ both $\F$ and $f$ are small and
(\ref{EQ:G-SCALE}) can be linearized. In this regime the physical solution to
(\ref{EQ:G-SCALE}) is
%
%
\be
\label{EQ:LARGEX}
\F \sim x^{d - 2 \lambda} e^{-x^{2}/2}.
\ee
These results are valid for arbitrary $n > 1$. Our focus, however, is  on
the $O(2)$ model where there
are known qualitative discrepancies with simulation and experimental data
\cite{LIU92b,MONDELLO90,NAGAYA95}. With this in
mind, we now  examine the small-$x$ behaviour of the scaling equation
(\ref{EQ:G-SCALE}) for the case $n=2$. For small-$x$ (\ref{EQ:G-SCALE})
admits the following general expansion for $f$:
%
%
\be
\label{EQ:EXPf}
f = 1 + f_{2} x^{2} \left[ 1 + \frac{K_{2}}{\ln x} \left( 1 + {\cal O}
\left[\frac{1}{\ln x} \right] \right) \right] + f_{4} x^{4} \left[
1 + \frac{K_{4}}{\ln x} \left( 1 + {\cal O} \left[\frac{1}{\ln x} \right]
\right) \right] + {\cal O} (x^{6}).
\ee
Non-analyticities in $f$ appear as a result of the non-zero $K_{2}$ and
$K_{4}$ coefficients multiplying factors of $1/\ln x$.

The coefficients of the expansion (\ref{EQ:EXPf}) can be
determined by examining
(\ref{EQ:G-SCALE}) order-by-order at small $x$. Balancing terms at
${\cal O} (\ln x)$ gives
%
%
\be
\label{EQ:F2DEF}
f_{2} =  - \frac{\pi}{4 \mu d}.
\ee
At ${\cal O}(1)$ one has
\be
K_{2} = - \frac{1}{d}.
\ee
The non-zero
$K_{2}$ coefficient is particularly important since it is responsible for
the divergence of the defect-defect correlation function at small $x$, as
one can see by examining (\ref{EQ:GDEFECTDEFECTCOR}) at small-$x$ where one
has, for $n=d=2$,
\be
\tilde{g} (x) = \frac{f_{2} K_{2}}{2 \pi^{2}} \frac{1}{x^{2} (\ln x)^{2} } +
 \cdots .
\ee
%
%
\section{The post-gaussian theory}
%
%
\subsection{Beyond the gaussian approximation}

Any scheme that claims to improve upon the gaussian approximation should
include the following elements: it should be
systematic, with the gaussian approximation being `lowest order', it should
be fairly easy computationally, at least in the early stages, and it should
be converging to the  results of experiment and simulation the farther one
goes in the scheme. We now present an approach which generalizes the
previous gaussian theory and satisfies the first two criteria completely,
and the third criterion in all areas save one.

Since any functional can be written as a sum over generalized Hermite
functional polynomials we write the probability distribution for the auxiliary
field in the form \cite{MAZENKO94a}
%
\be
P[\M] = P_{0}[\M] \sum^{\infty}_{J = 0} \mbox{ } \sum_{{i_{1},
 \cdots, i_{J}}} a_{J}(i_{1},\cdots,i_{J};1 \cdots J) H_{J}(i_{1},\cdots,
 i_{J};1 \cdots J)
\label{EQ:FULL P}
\ee
where the indices $i_{1} \cdots i_{J}$ each range from $1$ to $n$. Integration
over repeated spatial and temporal variables is assumed.
$P_{0}[\M]$ is the gaussian distribution. The $H_{J}$ are
the Hermite functional polynomials
%
\be
H_{J}(i_{1},\cdots,i_{J};1 \cdots J) = (-1)^{J} e^{K_{0}[\M]}
\frac{\delta^{J}}{\delta m_{i_{1}}(1) \cdots \delta m_{i_{J}}(J)}
e^{-K_{0}[\M]}
\label{EQ:HERMITE}
\ee
which form a complete orthogonal set, spanning a space containing the
 $O(n)$ symmetric
functionals \cite{MAZENKO94a}. The functions $a_{J}$,
along with $C_{0}$, are determined by the
symmetry of the problem and the series of constraints
%
\bea
\langle \vec{B}(1) \cdot \SG (2) \rangle & = & 0 \label{EQ:CST1}\\
\langle B_{i} (1) \sigma_{j} (2) \sigma_{k} (3) \sigma_{l}(4) \rangle & = & 0
 \label{EQ:CST3} \\ & \vdots & \nonumber
\eea
In the gaussian theory ($J = 0$) only the first constraint (\ref{EQ:CST1})
was necessary to completely determine the dynamics of $C_{0}$. As one does
computations at
 higher $J$ more of these
constraints are necessary to determine all the $a_{J}$. At each $J$ one has
a systematic approach to the problem and one  can, in principle,
calculate to any order in $J$. One expects that by
enforcing more constraints on $\vec{B}$ one will be satisfying the exact
equation of motion (\ref{EQ:SIGMAMOT}) more stringently. At the same time one
is
developing a more accurate expression for $P[\M]$ as more Hermite polynomials
are included. In this sense then the theory is expected to improve as one
calculates
to higher $J$. Finally, the use of Hermite polynomials allows one to
straightforwardly express
 post-gaussian averages in terms of easily computable gaussian averages.

We will work to order $J = 2$, which we call the first post-gaussian
approximation, so we will make the choices
%
\bea
\label{EQ:N2B}
a_{0} & = & 1 \\
a_{1}(i;1) & = & 0 \\
a_{2}(i,j;12) & = & \delta_{ij} A_{2}(12) \\
a_{J}(i_{1},\cdots, i_{J};1 \cdots J) & = & 0 \mbox{ for } J > 2.
\label{EQ:N2E}
\eea
Here $A_{2}(12)$ is a scalar function, symmetric in its arguments.
The condition on $a_{0}$ insures that the theory reduces to the correct
gaussian limit at lowest order and normalizes the probability distribution,
provided $P_{0}[\M]$ is normalized. The condition on $a_{1}$ reflects the fact
that there are no external fields and $P[\M] = P[-\M]$.
The choice for $a_{2}$ follows from the $O(n)$ symmetry, and considerations of
isotropy and homogeneity.
%
%
\subsection{General results for post-gaussian averages}

In order to calculate physical quantities like $C_{\psi}$ in the post-gaussian
approximation we must be able to express one- and two-point averages like
 $\langle \phi(1) \rangle$ and $\langle \phi (1) \chi (2) \rangle$,
where $\phi \mbox{ and } \chi$ are functions of $\M$, in terms of related
gaussian averages. The calculation for the case with no spatial gradients in
the average is presented below.  The important case where there are
spatial gradients in the average is slightly more involved, but the results for
the $O(n)$ model are straightforward generalizations of the results for the
scalar case, presented in \cite{MAZENKO94a}.
The two-point average $\langle \phi(1) \chi (2)\rangle$ can be related to
gaussian averages by using the definitions (\ref{EQ:FULL P}) and
(\ref{EQ:HERMITE}) and doing a few integrations by parts. The result is
\be
\label{EQ:PGSERIES}
\langle \phi(1) \chi(2) \rangle =
\sum_{J = 0}^{\infty} \mbox{ } \sum_{i_{1}, \cdots, i_{J}}
a_{J}(i_{1},\cdots,i_{J};\bar{1} \cdots \bar{J})
\left< \frac{\delta^{J} \phi(1)\chi(2)}
{\delta m_{i_{1}}(\bar{1}) \cdots \delta m_{i_{J}}(\bar{J})} \right>_{0}.
\ee
The barred quantities are integrated over in this expression, but they will
only contribute if they take the values 1 or 2.
In the first post-gaussian approximation (\ref{EQ:N2B}-\ref{EQ:N2E}) the
two-point average (\ref{EQ:PGSERIES}) is
%
\bea
\lefteqn{\langle \phi(1) \chi(2) \rangle = \langle
\phi(1) \chi(2) \rangle_{0} + A_{2}(11) \langle
\nabla_{m}^{2} \phi (1) \chi (2) \rangle_{0}}
\nonumber \\ & &
+ A_2(22) \langle \phi(1) \nabla_{m}^{2} \chi(2)
\rangle_{0} + 2 A_{2}(12)\langle \nabla_{m} \phi(1) \cdot
\nabla_{m} \chi(2) \rangle_{0},
\label{EQ:TWOPTN2}
\eea
with no integrations over the spatio-temporal variables.
If we now proceed under the assumption that $A_{2} (11) \mbox{ and }
A_{2} (22)$ are non-zero we will only be forced later to set them to
zero so that the long-distance behaviour of the theory is physical.
For notational convenience
we set these quantities to zero at the outset.
Equation (\ref{EQ:TWOPTN2}) can be further simplified by the introducing the
operator
%
\be
\hat{G}(12) = 1 + 2 A_{2} (12) \frac{\partial}{\partial C_{0} (12)}
\ee
since, for a general gaussian average $\langle \phi(1) \chi(2) \rangle_{0}$,
 the following identity holds:
\be
\langle \nabla_{m} \phi (1) \cdot \nabla_{m} \chi (2) \rangle_{0} =
\frac{\partial}{\partial C_{0}(12)} \langle \phi(1) \chi(2) \rangle_{0}.
\ee
The two-point post-gaussian average then can be compactly written
%
\be
\label{EQ:POSTOP}
\langle \phi(1) \chi(2) \rangle = \hat{G}(12) \mbox{ }
\langle \phi(1) \chi(2) \rangle_{0}.
\ee
The operator notation illuminates the close relation between gaussian and
post-gaussian averages in the first post-gaussian approximation.
By setting $\chi(2) = 1$ in (\ref{EQ:TWOPTN2}) one obtains the formula
%
\be
\langle \phi(1) \rangle = \langle \phi(1) \rangle_{0}
\ee
for the one-point average in the first post-gaussian approximation.
%
%
\subsection{$C_{\psi}$ and $C_{\psi^{2}}$ in the first post-gaussian
 approximation}

The post-gaussian analogue of (\ref{EQ:OPCOR2}) for the order-parameter
correlation function $C_{\psi}$ can be straightforwardly calculated using
(\ref{EQ:POSTOP}). We write
%
\bea
C_{\psi}(12) & = & \psi_{0}^{2} \mbox{ }
\hat{G}(12) \langle \hat{m} (1) \cdot \hat{m}(2) \rangle_{0} \nonumber \\
& = & \psi_{0}^{2} \F (12)
\eea
where now
\be
\F  = \frac{n}{2 \pi} B^{2} \left[ \frac{1}{2},
\frac{n+1}{2} \right] [ f F_{1} + 2 g (F_{1} + f F'_{1})].
\label{EQ:PGOPCOR}
\ee
The post-gaussian quantity $g$ is defined as
%
\be
g(12) \equiv   \frac{A_{2} (12)}{\bar{S_{0}}},
\ee
in analogy with the definition for $f$ (\ref{EQ:DEFNF}). Although $f$ retains
many features from the gaussian theory and $g$ carries much information
about the post-gaussian corrections, later we
will find that $f$ and $g$ influence each other strongly, and must be
determined together using the appropriate constraint equations.

Equation (\ref{EQ:POSTOP}) can also be used to determine the post-gaussian
$C_{\psi^{2}}$ from the gaussian $C_{\psi^{2}}$. We have
%
\be
C_{\psi^{2}} =  F_{2} + 2 g F'_{2} - 1 \mbox{ for } n \geq 2.
\label{EQ:PGCPSI2}
\ee

Unlike the situation in the gaussian theory, we now need to know the specific
forms for $f$ and $g$ in order to create a BBS plot. Knowledge of the
adjustable parameter in the length scale is, however, still unnecessary.
Since the forms for $f$ and $g$ are needed in the BBS plot, one hopes that the
shape of the BBS curve will now depend on spatial dimensionality.
We should note at this point that if we had maintained $A_{2}(11)
\neq 0$ we would now be forced to set it equal to zero so that both
$C_{\psi}$ and $C_{\psi^{2}}$ decay to zero at large distances, as is
expected physically.
%
%
\subsection{Equations of motion}

To determine the unknown functions $f$ and $g$ we will use the
constraints (\ref{EQ:CST1}) and (\ref{EQ:CST3}),
enforced to keep $\vec{B}$ small. For now, we will work with
$\T = 0$ to
make the post-gaussian analysis more transparent.

{}From our experience with the scalar case \cite{MAZENKO94a,MAZENKO94b}, we
know
that the constraint equation (\ref{EQ:CST1}) evaluated at a single space-time
point,
%
\be
\langle \vec{B}(1) \cdot \SG (1) \rangle = 0,
\label{EQ:SECONDCST}
\ee
contains information about the short-distance behaviour of the theory that
 allows
us later to simplify the constraint equations for $f$ and $g$.
Written in full,  (\ref{EQ:SECONDCST}) becomes
\be
\label{EQ:EQ1}
\frac{1}{2} \partial_{ t_{1}} \langle \sigma^{2}(1) \rangle -
\langle \SG(1) \cdot \nabla_{1}^{2} \SG(1) \rangle
+ \langle  \SG(1) \cdot  \nabla_{m}^{2} \SG(1) \rangle =
0.
\ee
Evaluating (\ref{EQ:EQ1}) to leading order, we have
%
\be
\label{EQ:D0DEF}
d_{0}^{(2)} = 1 \mbox{ for } n \ge 2
\ee
where
%
\bea
d_{0}^{(2)} & = & S_{0}^{(2)} + 2 A_{2}^{(2)} \\
S_{0}^{(2)} & \equiv & \lim_{1 \rightarrow 2} -\nabla^{2}_{1} C_{0} (12)
\label{EQ:ALTS0} \\
A_{2}^{(2)} & \equiv & \lim_{1 \rightarrow 2} -\nabla^{2}_{1} A_{2} (12).
\eea
The definition (\ref{EQ:ALTS0}) for $S_{0}^{(2)}$ is equivalent to
(\ref{EQ:DEFS02}) if one considers (\ref{EQ:DEFS02}) to be an average using
a gaussian probability distribution with the post-gaussian $C_{0}(12)$ as its
correlation function. The new constant $A_{2}^{(2)}$ will turn out to be
determined as
part of an eigenvalue problem. Since $S_{0}^{(2)}$ is a manifestly positive
quantity, (\ref{EQ:D0DEF}) also provides an important upper
bound on the eigenvalue $A_{2}^{(2)}$. We must have $A_{2}^{(2)} < 1/2$
for $n \geq 2$.

We now examine the constraint (\ref{EQ:CST1}) written out in full,
%
\be
\partial_{t_{1}} \langle \SG (1) \cdot \SG (2) \rangle - \nabla_{1}^{2}
\langle \SG (1) \cdot \SG (2) \rangle + \langle [\nabla_{m}^{2} \SG (1)]
\cdot \SG (2) \rangle = 0,
\label{EQ:PGMOT}
\ee
which in the gaussian case is the sole equation needed to determine the
dynamics of $C_{\psi}$. The evaluation of (\ref{EQ:PGMOT}) involves a
straightforward application of (\ref{EQ:POSTOP}) to the appropriate gaussian
averages. At late times the leading order result is
%
\be
\partial_{t_{1}} \F (12) -  \nabla_{1}^{2} \F (12)
- \frac{n}{2 \pi S_{0} (1)} B^{2} \left[ \frac{1}{2}, \frac{n+1}{2} \right]
[ (f + 2 g) ( F_{1} + f F'_{1} ) + 2 g f ( 2 F'_{1} + f F''_{1} ) ]= 0.
\label{EQ:PGMOTVER2}
\ee
For $t_{1} \gg t_{2}$  $\F$, $f$ and $g$ are small and
(\ref{EQ:PGMOTVER2}) again becomes a linear equation for $\F$. The relation
between $\lambda$ and $\mu$ is (\ref{EQ:GLAMBDA}), unchanged from the gaussian
theory, and the definition of the scaling length $L$ (\ref{EQ:GLENGTH}) is
retained.

As in the gaussian case we can write (\ref{EQ:PGMOTVER2}) as an equal-time
scaling equation
%
\be
{\vec x} \cdot \nabla_{x} \F + \nabla^{2}_{x} \F + \frac{n}{4 \mu}
 B^{2} \left[ \frac{1}{2},\frac{n+1}{2} \right] [ (f + 2 g)(F_{1} +
 f F'_{1}) + 2 g f (2 F'_{1} + f F''_{1})]
 = 0.
\label{EQ:PGSCALF}
\ee

Although the  functions $\F$,
$f$ and $g$ are related through  (\ref{EQ:PGOPCOR}) and (\ref{EQ:PGSCALF}),
the  additional
constraint equation (\ref{EQ:CST3}) is needed to complete the theory and
determine these functions separately. In (\ref{EQ:CST3}) we have a choice of
how to contract the indices $ijkl$ and $1234$. It is important to note that
the function $g$ entering the first post-gaussian theory is a two-point
quantity. Thus, unlike the usual case in perturbation theory, the first-order
corrections to the gaussian theory will not require us to treat the difficult
intricacies of four-point correlation functions. Therefore, in order to
determine $g$ we only need enforce (\ref{EQ:CST3}) contracted to a two-point
function.  A non-trivial
constraint is obtained by contracting the indices $ijkl$ in pairs. There then
remain two possible constraint equations:
%
\bea
\langle \vec{B}(1) \cdot \SG(1) \mbox{ } \SG(2) \cdot \SG(2) \rangle & = & 0
\label{EQ:THIRDCST} \\
\langle \vec{B}(1) \cdot \SG(2) \mbox{ }\SG(1) \cdot \SG(2) \rangle & = & 0.
\label{EQ:THIRDCST2}
\eea
Unless there exists some degeneracy one cannot, in the first post-gaussian
 approximation,
satisfy relations (\ref{EQ:THIRDCST}) and (\ref{EQ:THIRDCST2}) simultaneously.
Analysis of (\ref{EQ:THIRDCST2}) shows that for $n=2$ it
produces an equation in which the eigenvalue $A_{2}^{(2)}$ does not
appear \cite{WICKHAMU}. This is contrary to our expectation,
based on previous work
\cite{MAZENKO94a,MAZENKO94b}, that we have to solve a double eigenvalue problem
in the post-gaussian theory.
It is therefore clear that we should satisfy (\ref{EQ:THIRDCST}).

Since we are enforcing (\ref{EQ:SECONDCST}) we may rewrite
(\ref{EQ:THIRDCST}) as
\be
\langle \vec{B}(1) \cdot \SG(1) \Delta (2) \rangle =  0
\label{EQ:NICEFORM}
\ee
where
%
\be
\Delta (2) = \sigma^{2} (2) - \psi_{0}^{2}.
\ee
This computationally convenient form allows the calculation to proceed in
a way similar to that in \cite{MAZENKO94a}.
Written in full, (\ref{EQ:NICEFORM}) becomes
%
\be
\langle [\partial_{t_{1}} \SG(1)] \cdot \SG (1) \Delta (2) \rangle
- \langle [\nabla_{1}^{2} \SG(1)] \cdot \SG (1) \Delta (2) \rangle
+ \langle [\nabla_{m}^{2} \SG (1)] \cdot \SG (1) \Delta (2) \rangle  =  0.
\label{EQ:NICEFORM2}
\ee
To evaluate the post-gaussian averages in (\ref{EQ:NICEFORM2}) one must use
the formulae given in Section III.B and generalize the
results in \cite{MAZENKO94a} for post-gaussian averages containing spatial
gradients of a {\em scalar} field $m$ to the case of a vector field $\M$.
Two new gaussian averages specific to the $O(n)$ model
must be calculated. After some algebra and rearrangement,
(\ref{EQ:NICEFORM2}) reduces at late-times
to an equal-time scaling equation for $f$ and $g$.
For $n = 2$ we have, at leading order in $1/L$,
%
\be
4 \pi A_{2}^{(2)} g f + 4 \mu \nabla_{x} f \nabla_{x} g +
 \mu (\nabla_{x} f)^{2} ( 1 + 8 g f \gamma^{2}) = 0.
\label{EQ:SUPPN=2}
\ee
For $n > 2$ we have
%
\bea
\lefteqn{4 \pi A_{2}^{(2)} g f^{2} F'_{2} / \gamma^{2} + 4 \mu \nabla_{x} f
\nabla_{x} g \mbox{ } [ 2 f^{2} F_{2} - (n-2)(F_{2} -1)] +}
\nonumber \\ & &
\mu (\nabla_{x} f)^{2} \mbox{ } [ 2 f^{2} F_{2} - (n-2) (F_{2} -1) +
2 g ( 4 \gamma^{2} f^{3} F_{2} + 2 f^{2} F'_{2} -
\nonumber \\ & &
(n-2) ( F'_{2} + 2 (\gamma^{2} -2)(F_{2} -1 )/f))] = 0.
\label{EQ:SUPPNGT2}
\eea
Note that for $n=2$, $F_{2} = \gamma^{2}$ and  (\ref{EQ:SUPPNGT2}) reduces to
(\ref{EQ:SUPPN=2}).

These equations, together with (\ref{EQ:PGOPCOR}) and (\ref{EQ:PGSCALF}) form a
complete set of relations that will be used to  determine the functions
$\F$, $f$ and $g$. There are two unspecified constants in this set of
 equations
- $\mu$ and $A_{2}^{(2)}$. We thus have a non-linear eigenvalue problem
in which $\mu$, the eigenvalue familiar from the gaussian theory,
and the new eigenvalue $A_{2}^{(2)}$ are determined by connecting the
small- and large-$x$ behaviour of (\ref{EQ:PGSCALF}) and (\ref{EQ:SUPPNGT2}).

For large $x$ (\ref{EQ:PGSCALF}) reduces to a linear equation, as in the
gaussian case, and once again leads to (\ref{EQ:LARGEX}). The form for the
exponent $d - 2 \lambda$ in (\ref{EQ:LARGEX}) appears to be robust.
An examination of equation
(\ref{EQ:SUPPNGT2}) at large-$x$ yields
\be
\label{EQ:LGXFANDG}
f = - 4 g.
\ee
Up to now these results have been valid for $n > 1$. Now, in examining the
short scaled-distance properties, we will focus on the $n=2$ case where there
 are logarithmic corrections. The generalization of (\ref{EQ:EXPf}) is
%
\bea
\label{EQ:PGEXPf}
f(x) & =&  1 + f_{2} x^{2} \left[ 1 + \frac{K_{2}}{\ln x} \left( 1 + {\cal O}
\left[ \frac{1}{\ln x} \right] \right) \right] + f_{4} x^{4} \left[ 1 +
\frac{K_{4}}{\ln x} \left( 1 + {\cal O} \left[ \frac{1}{\ln x} \right] \right)
 \right] + {\cal O} (x^{6}) \\
\nonumber & & \\
\label{EQ:PGEXPg}
g(x) & = & g_{2} x^{2} \left[ 1 + \frac{L_{2}}{\ln x} \left( 1 + {\cal O}
\left[ \frac{1}{\ln x} \right] \right) \right]  + g_{4} x^{4} \left[ 1 +
\frac{L_{4}}{\ln x} \left( 1 + {\cal O} \left[ \frac{1}{\ln x} \right] \right)
 \right] + {\cal O} (x^{6}).
\eea
An examination of
(\ref{EQ:SUPPN=2}) yields, at  ${\cal O} (x^{2})$,
\be
\pi A_{2}^{(2)} g_{2} + \mu (f_{2})^{2} = 0.
\label{EQ:N=2,1}
\ee
At ${\cal O} (x^{2}/ \ln x)$ one has
\be
L_{2} = 2 K_{2}.
\label{EQ:N=2,2}
\ee
Examining (\ref{EQ:PGSCALF}) yields, at
 ${\cal O} (\ln x)$,
\be
f_{2} + 2 g_{2} = - \frac{\pi}{4 \mu d},
\label{EQ:N=2,3}
\ee
while at ${\cal O}(1)$, using (\ref{EQ:N=2,2}) one has
\be
K_{2} = - \frac{1}{2} + \frac{\pi}{4 \mu d (f_{2} + 4 g_{2})} \left[
\frac{2-d}{2d} - \frac{g_{2}}{f_{2}} \right].
\label{EQ:N=2,4}
\ee
Equations (\ref{EQ:N=2,1}) and (\ref{EQ:N=2,3}) determine $f_{2}$ and $g_{2}$
separately in terms of the eigenvalues $\mu$ and $A_{2}^{(2)}$
\bea
f_{2} & = & \frac{\pi}{4 \mu d} \left[
A_{2}^{(2)} d - \sqrt{A_{2}^{(2)} d (2 +
A_{2}^{(2)} d)} \mbox{ } \right] \label{EQ:N=2FORF2} \\
\nonumber & & \\
g_{2} & = & - \frac{\pi}{8 \mu d} \left[
1 + A_{2}^{(2)} d - \sqrt{A_{2}^{(2)} d
(2 + A_{2}^{(2)} d)} \mbox{ } \right]. \label{EQ:N=2FORG2}
\eea
We have assumed that $A_{2}^{(2)} > 0$ and taken the negative square root in
(\ref{EQ:N=2FORF2}) in order to render $f_{2}$ negative, as we expect
$f \leq 1$ physically. Equations (\ref{EQ:N=2,2}) and (\ref{EQ:N=2,4})
determine the corrections to the leading-order behaviour.
The small-$x$ expansion of the scaling form (\ref{EQ:PGCPSI2})
for $C_{\psi^{2}}$ is
%
\be
C_{\psi^{2}} (x) = \frac{2 g_{2} - f_{2}}{2 (f_{2})^{2}} \frac{1}{x^{2}} +
\frac{K_{2}}{2 f_{2}} \frac{1}{x^{2} \ln x} +
{\cal O} \left[ \frac{1}{(x \ln x)^{2}} \right].
\ee
Note that the condition for $C_{\psi^{2}}$ to be positive near the origin,
the behaviour expected physically, is $ 2 g_{2} - f_{2} > 0$
which is true for
%
\be
\label{EQ:ABOUND}
A_{2}^{(2)} > 1/4 d.
\ee
We will find, however, that $A_{2}^{(2)}$ does not
satisfy this lower bound, and the resulting negativity of
$C_{\psi^{2}}$ has to be interpreted carefully.
%
%
\subsection{Defect correlations}

The starting point for evaluating defect-defect correlations is to note
\cite{LIU92b} that the density for point defects (\ref{EQ:DEFDENSITY}) can be
rewritten in terms of the order-parameter field,
%
\be
\label{EQ:DEFOP}
\rho (1) = \delta[\OP(1)] \mbox{ det}[ \vec{\nabla}_{1} \OP(1)],
\ee
since the zeros of $\OP$ are the locations of the defects.
{}From (\ref{EQ:SIGEX}) and (\ref{EQ:A}) we see that $\OP \sim \M$ near the
defect cores so (\ref{EQ:DEFOP}) can be expressed in terms of the
auxiliary field
%
\be
\label{EQ:DEFAUX}
\rho (1) = \delta[\M(1)] \mbox{ det}[ \vec{\nabla}_{1} \M(1)].
\ee
This form is convenient for evaluating the equal-time
defect-defect correlations (\ref{EQ:DEFDENCOR}) which separate as indicated
in (\ref{EQ:POINTDEF}) into a piece representing the defect self-correlations,
%
\be
\label{EQ:DEFSELF}
n_{0}(t) = \langle \delta[\M(1)] | \mbox{det}[\vec{\nabla}_{1} \M(1)]| \rangle,
\ee
and a piece representing correlations between different defects
%
\be
\label{EQ:DEFOTHER}
g_{dd}({\bf r},t) = \langle \delta[\M(1)] \delta[\M(2)] \mbox{ det}[\vec{
\nabla}_{1}
\M(1)] \mbox{ det} [\vec{\nabla}_{2} \M(2)] \rangle.
\ee
We sketch the calculation of $g_{dd}({\bf r},t)$ for the post-gaussian theory
in the Appendix. We recover the scaling relation (\ref{EQ:DEFDEF})
with the post-gaussian scaling form
%
\be
\label{EQ:PGDEFDEF}
\tilde{g}(x) = n! \left( \frac{h}{x} \right)^{n-1} \left(
\frac{\partial h}{\partial x} + \frac{\partial \bar{h}}{\partial x} \right)
+ (n-1) n! \left(\frac{h}{x} \right)^{n-2} \frac{\bar{h}}{x} \frac{\partial h}
{\partial x}
\ee
where $h$ has the same definition it had in the gaussian theory
(\ref{EQ:HDEF}) and $\bar{h}$, which contains the new post-gaussian terms, is
given by
%
\be
\bar{h} = - \frac{\gamma}{\pi} \left(\frac{\partial g}{\partial x}
 + g f \gamma^{2} \frac{\partial f}{\partial x} \right).
\ee
The defect density $n_{0}(t)$ can be calculated directly by evaluating
the post-gaussian average in (\ref{EQ:DEFSELF}). The derivation is similar to
that given for the gaussian theory in \cite{LIU92b} and, as in the gaussian
theory, the absolute value of the determinant appearing in (\ref{EQ:DEFSELF})
is a complication that has to be carefully treated. Alternatively, we can
use the conservation law (\ref{EQ:N0GDD}) to compute $n_{0}(t)$, using the
post-gaussian formulae (\ref{EQ:PGDEFDEF}) for $\tilde{g}(x)$. Both approaches
lead to the same result:
%
\be
n_{0}(t) =  \frac{n!}{2^{n} \pi^{n/2} \Gamma(1 + n/2)}
            \left[ \frac{S_{0}^{(2)}}{n S_{0} (t)} \right]^{n/2}
            \left[1 + n \frac{A_{2}^{(2)}}{S_{0}^{(2)}} \right].
\ee
Using the small-$x$ expansions (\ref{EQ:PGEXPf}) and (\ref{EQ:PGEXPg})
for $f$ and $g$ we calculate the post-gaussian modification to the
behaviour of $\tilde{g}$ at small $x$ for $n=d=2$:
%
\be
\label{EQ:PGDEFSHT}
\tilde{g}(x) = \frac{( f_{2} + 4 g_{2} ) K_{2}}{2 \pi^{2}} \frac{1}{x^{2}
(\ln x)^{2}} + \cdots
\ee
Thus, as in the gaussian theory, leading order non-analyticities in the
small-$x$ behaviour of the post-gaussian $f$ and $g$ are responsible
for an unphysical divergence in $\tilde{g}(x)$ at small-$x$. We will
see that in the post-gaussian theory the divergence is weaker
than it was in the gaussian theory; however, through the inclusion
of fluctuations we can eliminate the divergence altogether.
%
%
\section{The Post-gaussian theory including fluctuations}
%
%
\subsection{Analysis of the $\SG$ degrees of freedom}

We will now show how the inclusion of fluctuations about the ordering field
$\SG$ can be used to eliminate the leading order non-analyticities in the
small-$x$ behaviour of $f$ and $g$, thereby rendering $\tilde{g}(0)$ finite.
This has been done successfully for the gaussian theory in \cite{MAZENKO96}.
Fluctuations influence the equation of motion for $\SG$ (\ref{EQ:SIGMAMOT})
through the non-zero $\T$ field (\ref{EQ:TFORM}).
The one-point equation (\ref{EQ:EQ1}), used to determine $d_{0}^{(2)}$
is modified by fluctuations to
%
\bea
\frac{1}{2} \partial_{t_{1}} \langle \sigma^{2} (1) \rangle  -
\langle \SG (1) \cdot \nabla_{1}^{2} \SG (1) \rangle
+ \langle \SG (1) \cdot \nabla_{m}^{2}
\SG (1) \rangle - \frac{\A}{L^{2}(t_{1})} \langle \sigma^{2} (1) \rangle  = 0.
\label{EQ:EQ1FLUCT}
\eea
The only difference between (\ref{EQ:EQ1FLUCT}) and (\ref{EQ:EQ1}) is the
last term, which is ${\cal O}(L^{-2})$. For $n=2$ the leading order
contribution to (\ref{EQ:EQ1FLUCT}) is ${\cal O}(\ln L / L^{2})$ so the
fluctuation term does not modify the result $d_{0}^{(2)} = 1$ for $n=2$ in
the post-gaussian theory. A similar thing happens in the gaussian theory where
the inclusion of fluctuations does not modify the relation $S_{0}^{(2)}=1$
for $n=2$ \cite{MAZENKO96}.

Fluctuations do, however, modify the two-point equation of motion
(\ref{EQ:PGMOT}) determining the order-parameter correlations and lead to
a new formula for $\lambda$:
%
\be
\label{EQ:PGFLUCTLAMBDA}
\lambda = d - \frac{\pi}{4 \mu} - \frac{\A}{2}.
\ee
At equal-times in the scaling regime  (\ref{EQ:PGMOT}),
including fluctuations, becomes
\be
\vec{x} \cdot \nabla_{x} \F + \nabla_{x}^{2} \F + \A \F +
\frac{n}{4 \mu} B^{2} \left[ \frac{1}{2}, \frac{n+1}{2} \right]
 [ (f + 2 g) (F_{1} + f F'_{1}) + 2 g f (2 F'_{1} + f F''_{1})] = 0.
\label{EQ:PGFSCALE}
\ee
The final equation of constraint (\ref{EQ:NICEFORM2}) is modified to
%
\be
\langle [\partial_{t_{1}} \SG(1)] \cdot \SG (1) \Delta (2) \rangle
- \langle [\nabla_{1}^{2} \SG(1)] \cdot \SG (1) \Delta (2) \rangle
-\frac{\A}{L^{2}(t_{1})} \langle \SG^{2} (1) \Delta (2) \rangle
+ \langle [\nabla_{m}^{2} \SG (1)] \cdot \SG (1) \Delta (2) \rangle = 0,
\ee
however a calculation for $n=2$ shows that at leading order in $1/L$,
this equation is unchanged from (\ref{EQ:SUPPN=2}),
even in the presence of fluctuations. Again, one has
%
\be
\label{EQ:3RDN=2FLUCT}
4 \pi A_{2}^{(2)} g f + 4 \mu \nabla_{x} f \nabla_{x} g + \mu
(\nabla_{x} f)^{2} ( 1 + 8 g f \gamma^{2}) = 0.
\ee
The large-$x$ behaviour of the theory with fluctuations is essentially
unchanged from the original post-gaussian theory. Equations (\ref{EQ:LARGEX})
and (\ref{EQ:LGXFANDG}) still hold except now $\lambda$ is given by
(\ref{EQ:PGFLUCTLAMBDA}).
The expansions (\ref{EQ:PGEXPf}) and (\ref{EQ:PGEXPg}) are used to
examine the small-$x$ behaviour of equations (\ref{EQ:PGFSCALE})
and (\ref{EQ:3RDN=2FLUCT}).
Once again $f_{2}$ and $g_{2}$ are given by
(\ref{EQ:N=2FORF2}) and (\ref{EQ:N=2FORG2}). The relation
(\ref{EQ:N=2,2}) still holds and from (\ref{EQ:PGFSCALE}) one has
%
\be
\A = g_{2} \left( 2 d + \frac{\pi}{2 \mu f_{2}} \right) - \frac{\pi}{2 \mu d}
 + 2 d K_{2} (f_{2} + 4 g_{2}).
\ee
At this point we insist that $f$ and $g$ are analytic for small-$x$
at ${\cal O}(x^{2})$ (i.e. $K_{2} = L_{2} = 0$).
This then fixes $\A$ in terms of $\mu$ and $A_{2}^{(2)}$:
%
\be
\label{EQ:W0DEFN}
\A = g_{2} \left( 2 d + \frac{\pi}{2 \mu f_{2}} \right) - \frac{\pi}{2 \mu d}.
\ee
The next order terms
are ${\cal O}(x^{4})$. From (\ref{EQ:PGFSCALE}) and (\ref{EQ:3RDN=2FLUCT})
we obtain two equations determining $f_{4}$ and $g_{4}$:
%
\be
(\pi A_{2}^{(2)} + 4 \mu f_{2} ) g_{4} + 4 \mu (f_{2}- g_{2}) f_{4} =
2 \mu g_{2} (f_{2})^{2} - \pi A_{2}^{(2)} g_{2} f_{2}
\ee
\be
f_{4} + 2 g_{4} =  \frac{3}{4} (f_{2})^{2} + 3 f_{2} g_{2} +
\frac{(f_{2} + 2 g_{2})}{4 (d+2)} \left[ \frac{\pi}{4 \mu} - 2 - \A \right].
\ee
Similarly, at the next order two relations can be derived for
$K_{4}$ and $L_{4}$, which are more involved.

By using $\A$ to eliminate the terms at
${\cal O}(x^{2}/\ln x)$ in $f$ and $g$ we have also managed to remove the
divergence in $\tilde{g} (x)$ at the origin. For small-$x$ we now have
%
\be
\tilde{g} (x) = - \frac{1}{\pi^{2}} \left( 3 (f_{4} + 2 g_{4}) -
\frac{(f_{2})^{2}}{2}  - 2 f_{2}g_{2} \right) + {\cal O} \left(\frac{1}{\ln x}
\right).
\ee
%
%
\subsection{Analysis of fluctuation correlations}

Correlations in the fluctuation field $\U$ are completely determined within
the theory. There are two types
of equal-time averages that are of interest to us. The first
describes cross-correlations between the $\SG$ and $\U$ fields and is defined
as
%
%
\be
\label{EQ:F1}
C_{u0}(12)= \langle \U ({\bf r}_{1},t)\cdot\SG ({\bf r}_{2},t)\rangle.
\ee
The second describes correlations of the fluctuation field with itself and is
given by
\be
\label{EQ:F2}
\delta_{ij} C_{uu}(12)=
\langle u_{i} ({\bf r}_{1},t) u_{j} ({\bf r}_{2},t)\rangle.
\ee
As we will see later these quantities are closely related in the scaling
regime. In \cite{MAZENKO96} it was shown how one can form equations of
motion for both $C_{u0}$ and $C_{uu}$ by using the equations of
motion (\ref{EQ:SIGMAMOT}) and (\ref{EQ:UONLY}) for $\SG$ and $\U$.
We can then use these equations to determine $C_{u0}$ and $C_{uu}$ explicitly
if we make the additional assumption that $\U$ is a gaussian field coupled to
the post-gaussian field $\M$. To effect this change we replace the gaussian
functional $e^{-K_{0}[\M]}$ in equations (\ref{EQ:FULL P}) and
(\ref{EQ:HERMITE}) by the gaussian functional $e^{-K_{0}[\M,\U]}$
which is quadratic in both $\M$ and $\U$. Post-gaussian averages over $\M$
are evaluated as before, while repeated use of the following identity
%
\be
\label{EQ:IDU}
\langle u_{i}(1) {\cal A} [\M,\U] \rangle = \int d3 \mbox{ }
C_{um}(13) \langle
\frac{\delta}
{\delta m_{i} (3)} {\cal A} [\M, \U] \rangle + \int d3
\mbox{ } C_{uu}(13) \langle \frac{\delta}
{\delta u_{i} (3)} {\cal A} [\M, \U] \rangle,
\ee
with
%
\be
\delta_{ij} C_{um} (12) = \langle u_{i} (1) m_{j} (2) \rangle,
\ee
allows averages over $\U$ to be expressed in terms $C_{uu}$, $C_{um}$
and post-gaussian averages over $\M$. We then evaluate the averages
over $\M$, which can be expressed in terms of $f$ and $g$ obtained previously.
Finally, we examine the equations of motion for the fluctuations
in the late-time scaling regime and extract the scaling functions.
The analysis closely follows that given in \cite{MAZENKO96} so
here we report only the final results for the post-gaussian theory.

In \cite{MAZENKO96} it was shown that, as a consequence of the definition
(\ref{EQ:TFORM}) for $\vec{\Theta}$,
we must have $ u \sim L^{-2}$ to leading order. We therefore write the scaling
relations
\be
C_{u0}(12)=\frac{\psi_{0}^{2}}{L^{2}}F_{u}(x)
\ee
and
\be
\label{EQ:FUUSCALE}
C_{uu}(12)=\frac{\psi_{0}^{2}}{L^{4}}F_{uu}(x).
\ee
The equations of motion developed in \cite{MAZENKO96} can be generalized to
the post-gaussian case and produce the following
relations between $F_{u}$ and $F_{uu}$:
%
\bea
\label{EQ:SCALEFORFU}
\lefteqn{
F_{u}(0) \left[\frac{1}{f} (1 - \sqrt{1 - f^{2}}) + 2 g \frac{1 -
\sqrt{1 - f^{2}}}{f^{2} \sqrt{1 - f^{2}}} \right]
} \nonumber \\ & &
+ F_{u} (x) \left[
\frac{\sqrt{1 - f^{2}}}{1 + \sqrt{1 - f^{2}} } - \frac{2 g f}{\sqrt{1 - f^{2}}
(1 + \sqrt{1 - f^{2}})^{2}} \right]
= - \frac{\A}{q_{0}^{2}} \F(x)
\eea
and
\be
\label{EQ:FUU}
F_{uu}(x) = - \frac{1}{q_{0}^{2}} \left[ \A + \frac{2 q_{0}^{2}}{\pi} F_{u}(0)
\right] F_{u}(x).
\ee
For $g=0$ (\ref{EQ:SCALEFORFU}) simplifies to the gaussian form found in
\cite{MAZENKO96}. The quantity $F_{u}(0)$ enters into these equations and
an analysis of (\ref{EQ:SCALEFORFU}) at $x=0$ gives
%
\be
\label{EQ:FUORIGIN}
F_{u}(0) = - \frac{\A}{q_{0}^{2}}.
\ee
We then find that
\be
\F_{uu}(0) = \frac{\A^{2}}{q_{0}^{4}} \left[ 1 - \frac{2}{\pi} \right] > 0,
\ee
which is a necessary condition for stability and is
expected from the definitions
(\ref{EQ:F2}) and (\ref{EQ:FUUSCALE}). Equations (\ref{EQ:SCALEFORFU}) and
(\ref{EQ:FUU}) explicitly show how correlations in the $\U$ field are slaved
 to those of the order parameter. We have demonstrated here that the theory is
consistent and that the fluctuations remain of ${\cal O}(L^{-2})$.
%
%
\section{Numerical Analysis of the Non-Linear Eigenvalue Problem}

\subsection{The post-gaussian theory without fluctuations}

The coupled non-linear equations (\ref{EQ:PGOPCOR}), (\ref{EQ:PGSCALF}) and
(\ref{EQ:SUPPN=2}) compose an eigenvalue problem for the eigenvalues $\mu$ and
$A_{2}^{(2)}$ which must be solved numerically. The eigenvalues are selected
by matching the small-$x$ behaviour, given by (\ref{EQ:PGEXPf}) and
(\ref{EQ:PGEXPg}), onto the behaviour at large $x$, equations (\ref{EQ:LARGEX})
and (\ref{EQ:LGXFANDG}). A fourth-order Runge-Kutta integrator is used to
integrate (\ref{EQ:PGSCALF}) and (\ref{EQ:SUPPN=2}) from near the origin
($x = 0.001$) into the large-$x$, asymptotic regime. Matching onto the proper
large-$x$ behaviour for $\F$ is the prime factor determining $\mu$; the value
of $A_{2}^{(2)}$ controls how well condition (\ref{EQ:LGXFANDG}) is satisfied.
The techniques used here are very similar to those used in previous studies
\cite{MAZENKO94a}.

We have examined the $O(2)$ model without fluctuations
 in two and three spatial dimensions.
Table \ref{TBL:EVALS} contains the results for the eigenvalues $\mu$ and
$A_{2}^{(2)}$. The upper bound $A_{2}^{(2)} < 1/2$ is satisfied both here, and
later when we include fluctuations. The order-parameter auto-correlation
exponent $\lambda$ can
be computed using (\ref{EQ:GLAMBDA}) once $\mu$ is known. The values for
$\lambda$ obtained from the post-gaussian theory are presented in Table
\ref{TBL:LAMBDA}, along with results from the gaussian theory \cite{MAZENKO96}.
The gaussian theory is in excellent agreement with simulations \cite{LEE95}
of the $O(2)$ model in two dimensions, which give the
value $\lambda = 1.171$.
The post-gaussian theory is in worse agreement with this simulation
result. We will see below that the
inclusion of fluctuations improves matters slightly.

While the post-gaussian theory decreases $\lambda$ significantly, the form of
the
order-parameter scaling function $\F$  changes only slightly
from the gaussian theory. The function $\F$ is plotted in Fig.
\ref{FIG:OPCOR} for the gaussian and post-gaussian theories in two dimensions.
The minor difference between the two theories is reassuring because the
gaussian theory is already in good agreement with simulations on this point
\cite{MAZENKO90}. The functional form of $f$ has the same qualitative features
as $\F$. The quantity $g$ is shown in Fig. \ref{FIG:G}. A key observation is
that the first post-gaussian correction measured by $g(x)$ is small for all
$x$.

The function $C_{\psi^{2}}$ (\ref{EQ:PGCPSI2}) measuring correlations in the
square of the order-parameter field can be calculated from $f$ and $g$.
The physical, positive divergence in $C_{\psi^{2}}$ at small-$x$, which is
seen in the gaussian
theory, is now rendered negative. This occurs in two and three dimensions.
One does not have this problem in the post-gaussian scalar theory
\cite{MAZENKO94a}.
Superficially, this unphysical result is a consequence of $A_{2}^{(2)}$ not
satisfying the lower bound (\ref{EQ:ABOUND}). More
careful consideration indicates that the root of the problem lies in the
method we chose to select $P[\M]$. Our truncation of the expansion of $P[\M]$
in Hermite functional polynomials (\ref{EQ:FULL P}) ignores terms in $P[\M]$
that are ${\cal O}(g^{2})$. To remedy this we use a corrected form for
$C_{\psi^{2}}$,
%
\be
\label{EQ:PGCPHI2COR}
C_{\psi^{2}} = F_{2} \left( 1 + g \frac{F'_{2}}{F_{2}} \right)^{2} - 1,
\ee
that differs from the previous form (\ref{EQ:PGCPSI2})
only at ${\cal O}(g^{2})$ and is manifestly positive near the origin. The BBS
plot of $1/(C_{\psi^{2}} + 1)$ against $1 - \F$ using the corrected form
(\ref{EQ:PGCPHI2COR}) is shown in Fig. \ref{FIG:BBS} for the $O(2)$ model in
two and three dimensions. The post-gaussian theory is in better agreement with
the simulation results \cite{BLUNDELL94} than the gaussian theory.
The two-dimensional post-gaussian results are in very good agreement with
simulation data for $1 - \F > 0.4$, which are intermediate to large distances,
$x > 1$. Unlike the gaussian theory, the shape of the BBS plot in the
post-gaussian theory depends on spatial dimensionality. On this point the
post-gaussian theory exhibits the same trends that are seen in the
simulations - as the dimensionality increases from two to three the BBS plot
approaches the gaussian result. These observations are strong evidence that
post-gaussian statistics provide a more accurate description of the
statistics governing the auxiliary field $\M$ than do gaussian statistics.

Fig. \ref{FIG:DEFECT} shows  the scaling form $\tilde{g}(x)$
(\ref{EQ:PGDEFDEF}) for the defect-defect correlations
for both the post-gaussian and the gaussian theories in two dimensions.
Simulation results \cite{MONDELLO90} are also shown. While there still is a
divergence in $\tilde{g}(x)$ at small-$x$ in the post-gaussian theory  it is
weaker than, and of opposite sign to, the divergence occuring in the gaussian
theory. The relative weakness of the divergence is a consequence of the small
value $K_{2} = 0.0018$ in the post-gaussian theory,
compared to $K_{2} = -0.5$ in the gaussian theory. We see that the
use of post-gaussian statistics does much to eliminate the the unphysical
divergence in $\tilde{g}(x)$, even in the absence of fluctuations.
%
%
\subsection{The post-gaussian theory including fluctuations}

The purpose of including fluctuations is four-fold.
First, we want to render the auxiliary field correlation function analytic for
small $x$.
Second, we would like to
completely
eliminate the unphysical divergence at small-$x$  in $\tilde{g}(x)$.
Third, we want to improve the agreement between the
value of $\lambda$ obtained from the post-gaussian theory and the value
seen in simulations. Fourth, we would like to see if the
inclusion of fluctuations increases $A_{2}^{(2)}$ so that
(\ref{EQ:ABOUND}) is satisfied. We choose $\T$ to have the form
(\ref{EQ:TFORM}).
In this eigenvalue problem $\A$ is selected using (\ref{EQ:W0DEFN}), which
guarantees that $K_{2} = L_{2} = 0$. We then solve the non-linear eigenvalue
problem
posed by (\ref{EQ:PGOPCOR}), (\ref{EQ:PGFSCALE}) and (\ref{EQ:3RDN=2FLUCT})
using the same methods we used above for the unmodified post-gaussian theory.
For $d=2$ the solution of the eigenvalue problem gives $\A = -0.2511$, while
for $d=3$ one has $\A = 0.1290$. The eigenvalues $\mu$ and $A_{2}^{(2)}$ are
given in Table \ref{TBL:EVALS}. For $d=2$ the eigenvalue
$A_{2}^{(2)}$ is increased; however, it still does not satisfy the bound
(\ref{EQ:ABOUND}). In three dimensions $A_{2}^{(2)}$ actually decreases
slightly.
Table \ref{TBL:LAMBDA} contains the values for $\lambda$.
The agreement between the value for $\lambda$ obtained from the post-gaussian
theory for $d=2$ and the value from simulations is slightly improved when
fluctuations are added. Adding fluctuations for $d=3$ actually decreases the
value for $\lambda$ slightly, moving the post-gaussian result farther away from
the gaussian result. We cannot comment on whether or not this represents an
improvement since, to our knowledge,  no simulation data for $\lambda$ for
the three-dimensional $O(2)$ model exists.

The scaling form $\F$ for order-parameter correlations is only slightly
modified from the form shown in Fig. \ref{FIG:OPCOR}. On this point, the
post-gaussian theory seems less susceptible to the perturbations introduced
by the fluctuations than the gaussian theory \cite{MAZENKO96}.
Since the bound (\ref{EQ:ABOUND}) is not satisfied we  must again use
(\ref{EQ:PGCPHI2COR}) for $C_{\psi^{2}}$ when creating the BBS plot.
The results are shown in Fig. \ref{FIG:BBS2}. It should be noted that
fluctuations do not modify the shape of the BBS plot in the gaussian theory
because the leading order dependence of $C_{\psi^{2}}$ and $\F$ on $f$ is
unaffected by $\U$, which is ${\cal O}(L^{-2})$. Again, we see that the
post-gaussian theory is in better qualitative agreement with the simulation
data than the gaussian theory. However, it appears that the addition of
fluctuations removes the dependency of the post-gaussian BBS plot on dimension.

The result for $\tilde{g}(x)$ in the fluctuation-modified theory
is shown in Fig. \ref{FIG:DEFECT}. We see that the inclusion of fluctuations
not only eliminates the divergence of $\tilde{g}(x)$ at the origin, but that
$\tilde{g}(x)$ is in better agreement with the simulation results down to
smaller values of $x$ than it was in the unmodified post-gaussian theory.
When compared to the gaussian theory with fluctuations \cite{MAZENKO96} we
see that the addition of post-gaussian corrections has the desired effect of
reducing the magnitude of $\tilde{g}(0)$ and thus producing better agreement
with simulation results.
%
%
\section{Discussion}

The main achievement of this paper is the extension of
the gaussian theory for the $O(2)$ model to allow
for non-gaussian statistics. In this sense the theory is unique. It is not
trivial that the scaling properties of the gaussian theory are retained.
We have demonstrated that when the
structural form of $P[\M]$ is modified to include a post-gaussian part we can
still solve the (now double) eigenvalue problem and find the fixed-point
solution for $P[\M]$.
Thus we see a structure emerging, related to the
nature of the fixed-point, that encourages us to believe that we can
systematically change the structural form for $P[\M]$ {\em via}
(\ref{EQ:FULL P}) and obtain improved results for the scaled quantities.
Although the new function $g$ is small, the approach we
are attempting here is non-perturbative.

In all areas examined, save the determination of the exponent $\lambda$,
the first post-gaussian approximation is in better agreement with the
simulation results than the gaussian theory. Scaling results that are already
in good agreement with simulations, such as $\F$ and the large scaled-distance
behaviour of $\tilde{g}(x)$ are essentially unmodified by the addition of
post-gaussian corrections. On the other hand in the BBS plot (Fig.
\ref{FIG:BBS}), where there exists a large discrepancy between simulations
\cite{BLUNDELL94} and the gaussian theory, the addition of post-gaussian terms
qualitatively improves matters. Unlike the gaussian theory, the BBS plot
in the post-gaussian theory without fluctuations has a dependence on spatial
dimensionality which shows the same trends seen in the simulations.
In addition, the first post-gaussian approximation nearly removes the
divergence in $\tilde{g}(x)$ at small $x$.

The development here suggests some avenues for improving the post-gaussian
treatment. For now we have used the {\em ad hoc} formula
(\ref{EQ:PGCPHI2COR}), which is correct to ${\cal O}(g)$, to obtain
meaningful results for the BBS plot.  In order to properly treat these
corrections at ${\cal O}(g^{2})$ we should go to the next order in the
post-gaussian approximation sequence. In the current approximation the value
for $\lambda$ is decreased from the gaussian result, and one hopes
that the systematic inclusion of terms at ${\cal O}(g^{2})$ will raise
$\lambda$ and provide better agreement with simulations.
It is apparent that within this approximation scheme the value for
$\lambda$ is converging slower than the results for the scaling functions,
and may even be experiencing some type of oscillatory behaviour.

It is also a non-trivial matter that we were able to incorporate fluctuations
into the post-gaussian theory in a consistent manner. We have
been able to successfully render both
$f$ and $g$ more analytic and thus eliminate
the divergence in $\tilde{g}(x)$. The addition of post-gaussian
terms to the fluctuation-modified gaussian theory has brought the value for
$\tilde{g}(0)$ into better agreement with simulations and thus generally
improved the agreement at small-$x$. The fluctuations also slightly
increase the value for $\lambda$.  We would also like to go beyond the
the simplest post-gaussian fluctuation theory, represented by
(\ref{EQ:TFORM}). To do this it would seem that one needs to consider
$\U$ as a post-gaussian field, coupled to the post-gaussian $\M$
\cite{WICKHAMU}. Another interesting effect of adding fluctuations to
the post-gaussian theory is the elimination of the dependence of the
BBS plot on spatial dimensionality. Whether this is by accident,
or is due to some deeper structure in the theory is presently unknown.
It would be interesting to see if this
effect remains if one includes ${\cal O} (g^{2})$ terms in $P[\M]$.

An obvious next step would be to examine the $O(3)$ model using post-gaussian
statistics. The $O(3)$ model is interesting to study because of it's
application to ferromagnetic materials and the role it plays in the
description of monopole defects in nematic liquid crystals and cosmology
\cite{PARGELLIS91}.
There are some interesting new aspects associated with higher $n$
\cite{WICKHAMU}.

In examining the $O(2)$ model we have shown how the post-gaussian theory
generalizes to a situation with continuous symmetry and solves some of the
problems specific to the $n=2$ case. In developing the post-gaussian
theory as a systematic and calculable extension of the gaussian theory we
have also established the role that the gaussian theory plays as a
zeroth-order approximation to the true statistics.
There are many aspects of this process
that suggest a deeper underlying structure in the theory, including the
interesting interactions between the post-gaussian corrections and the
fluctuations.
%
%
\acknowledgements
R.A.W. gratefully acknowledges support from the NSERC of Canada.
This work was supported in part by the MRSEC Program of the National Science
Foundation under Award Number DMR-9400379.
%
%
\appendix
\section*{}

In this appendix we outline the method used to evaluate the post-gaussian
average occuring in the defect-defect correlation function $g_{dd}({\bf r},t)$
(\ref{EQ:DEFOTHER}). We have
%
\bea
g_{dd}
({\bf r},t) & = & \langle \delta[\M(1)] \delta[\M(2)] \mbox{ det}[\vec{\nabla}
_{1} \M (1)] \mbox{ det}[ \vec{\nabla}_{2} \M (2) ] \rangle
\nonumber \\
& = & \epsilon_{i_{1} \cdots i_{n}} \epsilon_{j_{1} \cdots j_{n}}
\langle D_{i_{1}}(1) \cdots D_{i_{n}}(1) D_{j_{1}}(2) \cdots D_{j_{n}}(2)
\rangle,
\label{EQ:DEFCOREX}
\eea
with a sum over repeated indices and
%
\be
D_{i_{q}}(1) = \delta[m_{q}(1)] \frac{\partial m_{q} (1)}{\partial r_{1i_{q}}}.
\ee
Using the definitions (\ref{EQ:N2B}-\ref{EQ:N2E}) and (\ref{EQ:PGSERIES})
the post-gaussian average in (\ref{EQ:DEFCOREX}) is
%
\bea
\lefteqn{
\langle D_{i_{1}}(1) \cdots D_{i_{n}}(1) D_{j_{1}}(2) \cdots D_{j_{n}}(2)
\rangle = \langle D_{i_{1}}(1) \cdots D_{i_{n}}(1) D_{j_{1}}(2) \cdots
D_{j_{n}}(2) \rangle_{0}
} \nonumber \\ & &
  \nonumber \\ & &
+ \sum_{k=1}^{n} \int d\bar{1} d\bar{2} A_{2}(\bar{1}
\bar{2}) \left<  \frac{\delta^{2}}{\delta m_{k}(\bar{1}) \delta m_{k}
(\bar{2})} [ D_{i_{1}}(1) \cdots D_{i_{n}}(1) D_{j_{1}}(2) \cdots
D_{j_{n}}(2)] \right>_{0}.
\label{EQ:FULLPGEXP}
\eea
The first term on the right-hand side of (\ref{EQ:FULLPGEXP}) is the original
gaussian average, and was computed in \cite{LIU92b}.
We focus on the term involving $A_{2}$, which factors into the following
 product of $n$ averages over the gaussian distributed {\em scalar} $m_{i}$'s:
%
\be
\sum_{k=1}^{n} \int d\bar{1} d\bar{2} A_{2}(\bar{1} \bar{2}) \left<
\frac{\delta^{2}}{\delta m_{k}(\bar{1}) \delta m_{k}(\bar{2})}
[D_{i_{k}} (1)  D_{j_{k}} (2)] \right>_{0} \mbox{ }\prod_{q = 1, q \neq k}^{n}
\langle D_{i_{q}}(1) D_{j_{q}}(2) \rangle_{0}.
\label{EQ:FACTOREDPG}
\ee
{}From \cite{LIU92b} we know that
%
\be
\langle D_{i_{q}} (1)  D_{j_{q}} (2) \rangle_{0} = \frac{h}{r}
\delta_{i_{q}j_{q}} + \left[ \frac{\partial h}{\partial r} - \frac{h}{r}
\right] \hat{r}_{i_{q}} \hat{r}_{j_{q}}
\ee
with
%
\be
h = - \frac{\gamma}{2 \pi} \frac{\partial f}{\partial r}.
\ee
In (\ref{EQ:FACTOREDPG}) the factor involving the integral is evaluated by
operating with the functional derivatives, integrating by parts, applying the
resulting delta functions and calculating the remaining gaussian
integrals. We find
%
\be
\int d\bar{1} d\bar{2} A_{2}(\bar{1} \bar{2}) \left<
\frac{\delta^{2}}{\delta m_{k}(\bar{1}) \delta m_{k}(\bar{2})}
 [D_{i_{k}} (1)  D_{j_{k}}(2)] \right>_{0}
= \frac{\bar{h}}{r} \delta_{i_{k}j_{k}} + \left[
\frac{\partial \bar{h}}{\partial r} - \frac{\bar{h}}{r} \right]
\hat{r}_{i_{k}} \hat{r}_{j_{k}}.
\ee
with
%
\be
\bar{h} = - \frac{\gamma}{\pi} \left(\frac{\partial g}{\partial r}
 + g f \gamma^{2} \frac{\partial f}{\partial r} \right).
\ee
Thus the complete expression for $g_{dd}$ is
\be
g_{dd}({\bf r},t) = n! \left( \frac{h}{r} \right)^{n-1}
\frac{\partial h}{\partial r} + \Delta
\ee
with the post-gaussian terms contained in
%
\be
\Delta = \sum_{k=1}^{n} \epsilon_{i_{1} \cdots i_{n}}
\epsilon_{j_{1} \cdots j_{n}} \left( \frac{\bar{h}}{r} \delta_{i_{k}j_{k}} +
 \left[ \frac{\partial \bar{h}}{\partial r} - \frac{\bar{h}}{r} \right]
\hat{r}_{i_{k}} \hat{r}_{j_{k}} \right)
\prod_{q = 1, q \neq k}^{n} \left( \frac{h}{r}
\delta_{i_{q}j_{q}} +
 \left[ \frac{\partial h}{\partial r} - \frac{h}{r} \right]
\hat{r}_{i_{q}} \hat{r}_{j_{q}} \right)
\ee
Due to symmetry, the only non-zero contributions to $\Delta$ come from
terms containing either zero or exactly two factors of $\hat{r}$.
Mindful of this, we evaluate $\Delta$ and obtain
\be
\Delta = n! \left(\frac{h}{r} \right)^{n-2} \left[
\frac{\partial \bar{h}}{\partial r} \frac{h}{r} + (n-1) \frac{\bar{h}}{r}
\frac{\partial h}{\partial r} \right].
\ee
The post-gaussian formula for $g_{dd}$ is then
%
\be
g_{dd} ({\bf r},t) = n! \left( \frac{h}{r} \right)^{n-1}  \left(
 \frac{\partial h}
{\partial r} + \frac{\partial \bar{h}}{\partial r} \right) + (n-1) n!
\left( \frac{h}{r} \right)^{n-2} \frac{\bar{h}}{r}
\frac{\partial h}{\partial r}.
\ee
%
%

%
%
\begin{table}
\caption{Values for the eigenvuales $\mu$ (top section) and $A_{2}^{(2)}$
(bottom section) for the post-gaussian theory, both with and without
fluctuations.}
\label{TBL:EVALS}
\begin{tabular}{ccccc}
        & $ \A = 0$   &  $\A \neq 0$ \\
\tableline
$d=2$   & 0.8380	   &  0.7597 	            \\
$d=3$   & 0.5326	   &  0.5525  		    \\
\tableline
$d=2$   & 0.05861          &  0.08092		    \\
$d=3$   & 0.06364          &  0.05392
\end{tabular}
\end{table}
%
%
\begin{table}
\caption{Auto-correlation exponent $\lambda$ from the gaussian
 theory \protect\cite{MAZENKO96} (top section) and from
the post-gaussian theory (bottom section), both with and without fluctuations.}
\label{TBL:LAMBDA}
%
\vspace{.5 in}
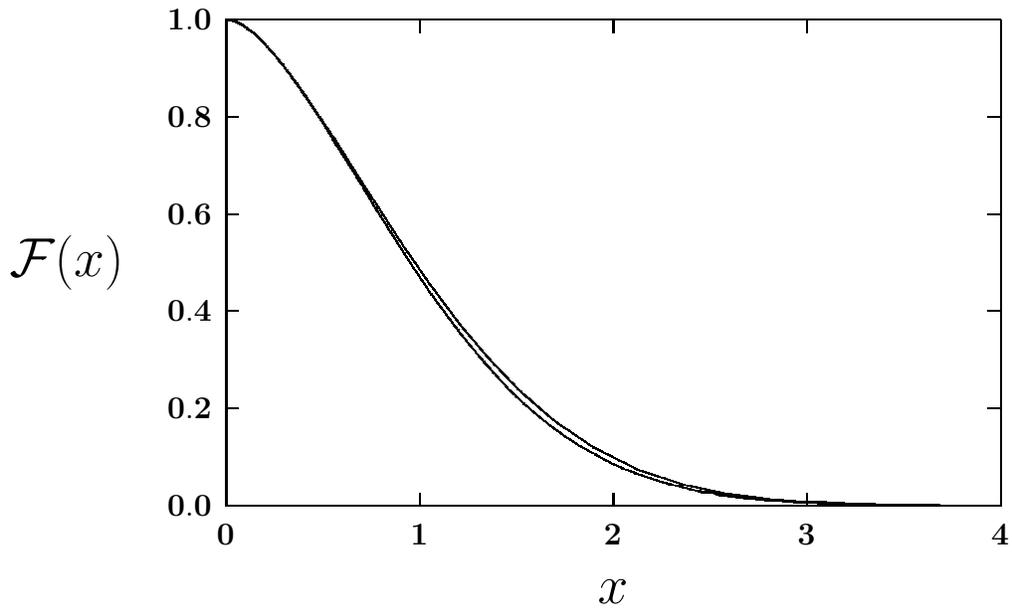
\caption{Scaling form $\F(x)$ for the order-parameter correlation function in
two dimensions. At $x=2$ the upper curve is the post-gaussian theory without
fluctuations and the lower curve is the gaussian theory without fluctuations.}
\label{FIG:OPCOR}
\end{figure}
\pagebreak
%
%
\begin{figure}
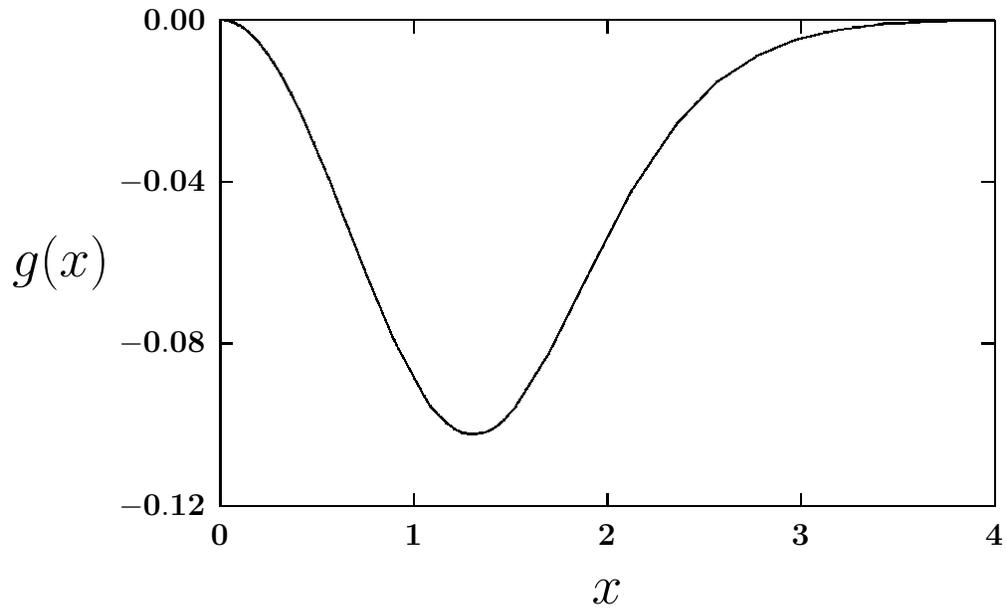

\setlength{\unitlength}{0.240900pt}
\ifx\plotpoint\undefined\newsavebox{\plotpoint}\fi
\sbox{\plotpoint}{\rule[-0.200pt]{0.400pt}{0.400pt}}%
%
\vspace{.5 in}
\caption{The auxiliary function $g(x)$ for the post-gaussian theory without
fluctuations in two dimensions.}
\label{FIG:G}
\end{figure}
\pagebreak
%
%
\begin{figure}
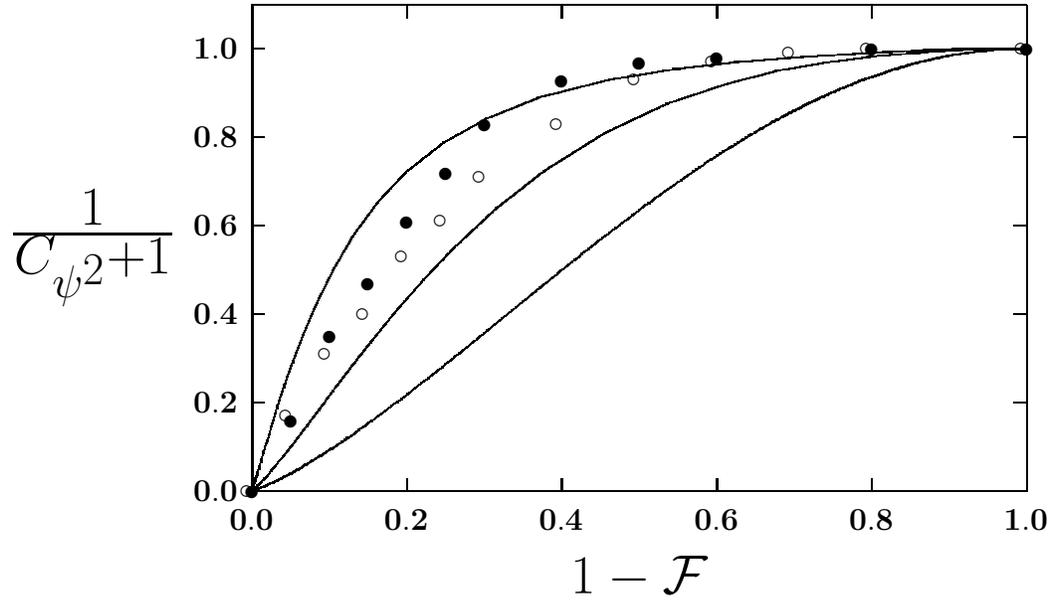

\setlength{\unitlength}{0.240900pt}
\ifx\plotpoint\undefined\newsavebox{\plotpoint}\fi
\sbox{\plotpoint}{\rule[-0.200pt]{0.400pt}{0.400pt}}%
%
\vspace{.5 in}
\caption{BBS plot for the gaussian and post-gaussian theories without
fluctuations. At
$1 - \F = 0.2$ the upper solid curve is the post-gaussian result for $d=2$,
the middle curve is for $d=3$ and the lower curve is the gaussian result.
The solid circles are the simulation data for $d=2$ and the open circles are
the simulation data for $d=3$ \protect\cite{BLUNDELL94}.}
\label{FIG:BBS}
\end{figure}
\pagebreak
%
%
\begin{figure}
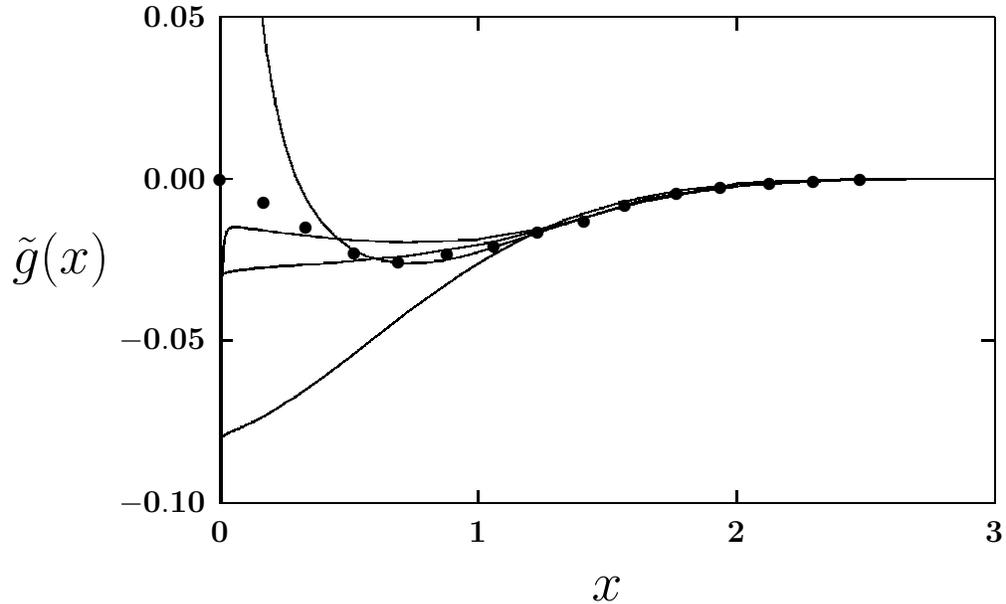

\setlength{\unitlength}{0.240900pt}
\ifx\plotpoint\undefined\newsavebox{\plotpoint}\fi
\sbox{\plotpoint}{\rule[-0.200pt]{0.400pt}{0.400pt}}%

\vspace{.5 in}
\caption{Scaling function $\tilde{g}(x)$ for the defect-defect correlations
in two dimensions. At $x = 0.1$, from bottom to top, the solid curves
represent: the gaussian theory with $\A \neq 0 $ \protect\cite{MAZENKO96}; the
post-gaussian theory with $\A \neq 0$; the post-gaussian
theory without fluctuations (diverging negatively); the gaussian theory
without fluctuations (diverging positively) \protect\cite{LIU92b}. The dots
 represent
the simulation data \protect\cite{MONDELLO90}.}
\label{FIG:DEFECT}
\end{figure}
\pagebreak
%
%
\begin{figure}
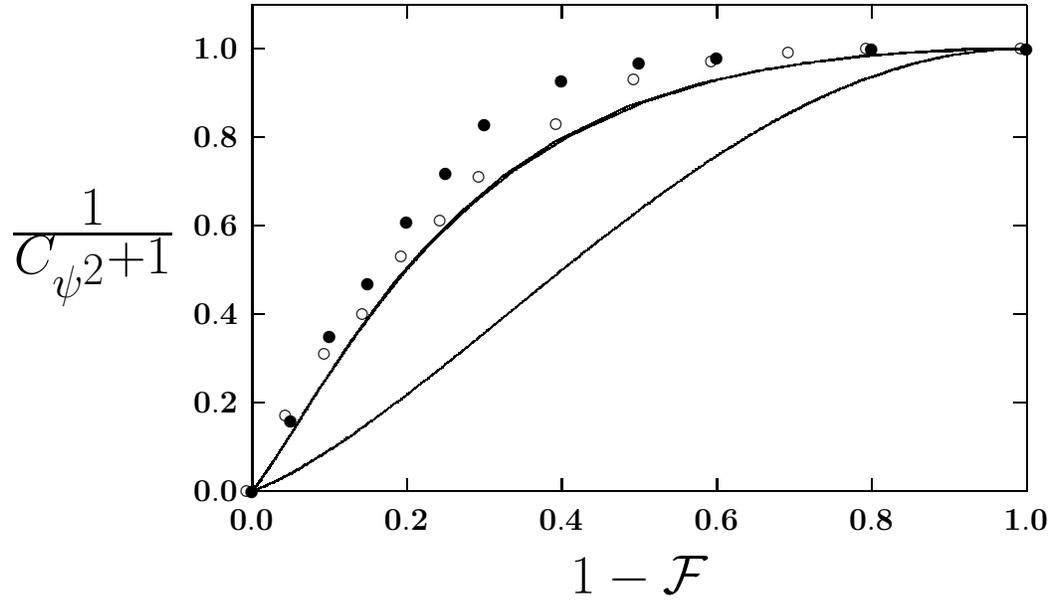

\setlength{\unitlength}{0.240900pt}
\ifx\plotpoint\undefined\newsavebox{\plotpoint}\fi
\sbox{\plotpoint}{\rule[-0.200pt]{0.400pt}{0.400pt}}%

\vspace{.5 in}
\caption{BBS plot for the post-gaussian theory including fluctuations. At
$1 - \F = 0.5$ the lower solid curve is the gaussian theory and
the upper two curves, which are virtually indistinguishable,  are the
post-gaussian theory with $\A \neq 0$ for $d=2$ and $d=3$.
The solid circles are the
simulation data for $d=2$ and the open circles are the simulation data for
$d=3$ \protect\cite{BLUNDELL94}.}
\label{FIG:BBS2}
\end{figure}

\begin{thebibliography}{99}
\bibitem{PREFACTOR} This is not because one cannot determine the pre-factor
                    theoretically but because of the difficulty of
                    discrete lattices used in simulations.
\bibitem{POTENTIAL} Typically, $V$ is taken to have the form:
                    $V = \frac{1}{4} (\psi_{0}^{2} - \psi^{2})^{2}$.
\bibitem{MASS} There is a mass term in the equation for $\U_{T}$, but it is
               ${\cal O} (L^{-2})$.
\bibitem{MU} This relation between $L^{2}(t)$ and $S_{0}(t)$, which introduces
             the constant $\mu$ is the same relation as in \cite{LIU92a}. In
             other work \cite{MAZENKO90} one defines $\mu$ through the
             relations $L \dot{L} = \mu$, $L^{2} = \pi S_{0}$. This approach is
             physically appealing because $\mu$ then determines the pre-factor
	     to the power-law growth: $L = \sqrt{2 \mu} t^{1/2}$.
             Either approach, however, is valid.
\end{thebibliography}
\end{document}